\documentclass[useAMS, usenatbib]{mn2e}
\usepackage{times}
\usepackage{amsmath}
\usepackage{amssymb}
\usepackage{graphicx}
\usepackage{subfigure}
\usepackage[letterpaper,margin=0.65in]{geometry}
\usepackage{natbib}
\usepackage{hyperref}
\usepackage{enumitem}
\usepackage[usenames]{color}

\title{The Dynamical Origin of Early-Type Dwarfs in Galaxy Clusters: A Theoretical Investigation}
\author[R. Vijayaraghavan, J.S. Gallagher, and P.M. Ricker]
      {Rukmani Vijayaraghavan$^1$\thanks{E-mail: vijayar2@illinois.edu}, John S. Gallagher, III$^2$, and Paul M. Ricker$^1$\\
        $^1$Department of Astronomy, University of Illinois at Urbana-Champaign, 1002 W. Green Street, Urbana, IL 61801, USA \\
        $^2$Department of Astronomy, University of Wisconsin-Madison, 475 N. Charter St., Madison, WI 53706, USA}

\begin{document}

\date{\today}
\maketitle

\begin{abstract}
  Observations of early-type dwarf galaxies in clusters often show that cluster dwarf members have significantly higher velocities and less symmetric distributions than cluster giant ellipticals, suggesting that these dwarfs are recently accreted galaxies, possibly from an infalling group. We use a series of $N$-body simulations, exploring a parameter space of groups falling into clusters, to study the observed velocity distributions of the infall components along various lines of sight. We show that, as viewed along a line of sight parallel to the group's infall direction, there is a significant peculiar velocity boost during the pericentric passage of the group, and an increase in velocity dispersion that persists for many Gyr after the merger. The remnants of the infalling group, however, do not form a spatially distinct system -- consistent with recent observations of dwarf galaxies in the Virgo and Fornax clusters. This velocity signature is completely absent when viewed along a direction perpendicular to the merger. Additionally, the phase-space distribution of radial velocity along the infall direction versus cluster-centric radius reveals the separate dynamical evolution of the group's central core and outer halo, including the presence of infalling remnants outside the escape velocity envelope of the system. The distinct signature in velocity space of an infalling group's galaxies can therefore prove important in understanding the dynamical history of clusters and their dwarfs. Our results suggest that dwarf galaxies, being insensitive to dynamical friction, are excellent probes of their host clusters' dynamical histories. 
\end{abstract}


\section{Introduction}
\label{sec:intro}

Galaxy clusters are the youngest and most massive gravitationally bound objects in the Universe. In the cold dark matter paradigm of structure formation, they grow by accreting smaller galaxies and groups of galaxies. Since the dynamical timescale in clusters is typically on the order of a Gyr, a significant fraction of clusters exhibit substructure in the spatial and velocity distributions of their galaxies (\citealt{deVaucouleurs61}, \citealt{Geller82}, \citealt{Dressler88}). In principle, then, it should be possible to quantify the dynamical states of clusters through measurements of the velocity clustering of cluster galaxies.  The morphology and color dependence of this clustering should provide information about cluster assembly and the processes that transform cluster galaxies.

The principal difficulty in carrying out such measurements lies in obtaining sufficiently many cluster galaxy spectra. Hence cluster velocity substructure analyses had to wait for technology to enable the rapid determination of galaxy spectra. \citet{deVaucouleurs61}, in a pioneering work based on 212 Virgo galaxies (79 of which then had known radial velocities measured with the Palomar spectrograph), showed that the Virgo Cluster was constituted of at least two `clouds': a concentration of elliptical and lenticular galaxies, with a velocity dispersion of $\sim 550$  km s$^{-1}$, and a second concentrated cloud of primarily spiral and irregular galaxies with a velocity dispersion of $\sim 750$  km s$^{-1}$. \citet{Binggeli87}, in a later study enabled by higher resolution observations of fainter Virgo members in the deep Las Campanas survey, studied 1277 Virgo galaxies, of which 572 had known radial velocities. They found evidence for substructure in Virgo centered on M87 and M49. They also found that late-type galaxies in Virgo have a significantly larger velocity dispersion ($890$ km s$^{-1}$) than early-type galaxies ($570$ km s$^{-1}$), and are less centrally concentrated than early-types, suggesting that late-type galaxies are currently infalling. \citet{Beers82}, in another early study, calculated the line-of-sight velocity dispersions of two individual subclusters in the cluster Abell 98 using radial velocity measurements of 13 galaxies. \citet{Dressler88} using radial velocity measurements of galaxies in 15 clusters, estimated that 30-40 \% of clusters have significant substructure. \citet{Colless96}, using redshift measurements of 552 Coma Cluster galaxies (including 243 new measurements with the KPNO Hydra spectrograph), showed that the Coma Cluster is in the process of merging with at least two subclusters: the NGC 4839 group, which has a relative velocity of $\sim 1700$ km s$^{-1}$ with respect to the main cluster and a physical separation of $\sim 0.8 \, h^{-1}$ Mpc, and a subcluster centered on NGC 4889 with a relative velocity of $\sim 1200$ km s$^{-1}$. 

Obtaining sufficent spectra for substructure analyses also requires that enough galaxies be present and detectable. Dwarf galaxies are the most common type of galaxies in clusters, and as evidenced by the steeper luminosity functions in clusters compared to the field, clusters have a higher dwarf-to-giant galaxy ratio than the field (e.g., \citealt{Binggeli88}, \citealt{Bernstein95}, \citealt{dePropris95}, \citealt{Lobo97}, \citealt{Smith97}, \citealt{Secker97}, \citealt{Milne07}, \citealt{Lu09}, \citealt{deFilippis11}). The enhanced dwarf-to-giant ratio in clusters is most likely a consequence of efficient tidal stripping and harassment of galaxies in dense environments (\citealt{Moore96}, \citealt{Moore99}, \citealt{Gnedin03b}, \citealt{Gnedin03},  \citealt{Villalobos12}, \citealt{VR13}, \citealt{Villalobos14}). Thus among cluster galaxies it is the dwarfs that should provide the best tracers of overall cluster dynamics and the extent to which cluster galaxies have been transformed in dense environments.

In one of the earliest studies of cluster dwarf dynamics, \citet{Binggeli93} found that the velocity distribution of dwarf ellipticals in the core of Virgo is highly asymmetric, suggesting the presence of a merging subcluster in the core region centered on M86 in addition to the main cluster centered on M87. \citet{Conselice01}, in a more detailed study based on the radial velocities of 141 dE + dS0 galaxies (dwarf ellipticals and spheroidals) in Virgo showed that early-type dwarfs in Virgo resemble the expected remnants of infalling field galaxies. The velocity dispersion ratio of early-type dwarfs to giants is consistent with that of infalling to virialized populations, and dE galaxies are not spatially concentrated, unlike giant ellipticals. \citet{Lisker09} subdivided the population of Virgo dE's into fast- and slow-moving dE's, and found that fast-moving dE's are more likely to be on radial orbits and have flattened shapes, while slow-moving dE's are likely on circular orbits and have rounder shapes. This is consistent with the two populations being recently accreted and older, respectively. The dynamics of dE's in other clusters is similarly indicative of cluster formation and galaxy transformation histories. \citet{Drinkwater01} analysed the radial velocities of 108 galaxies in the Fornax Cluster, and found that the velocity dispersion of the  dwarf galaxy population is $\sim 1.4$  times larger than the giant galaxies' velocity dispersion, consistent with the dwarfs being an infall population.  

After obtaining enough galaxy spectra, the second major limitation on velocity substructure analyses is the fact that we can only measure the line-of-sight component of velocity. The detectability of velocity substructure is therefore diminished for unfavorable geometries (e.g., mergers in the plane of the sky). The practical impact of this limitation can be investigated using $N$-body simulations. For example, \citet{Pinkney96}, using a non-cosmological approach, showed that the sensitivity of substructure detection increased with the addition of velocity information, particularly with head-on mergers, and that mergers skewed velocity distributions. On average, 20 -- 30 of the 36 cases they studied were detected at less than 10\% significance (compared to a null hypothesis of no substructure), and 15 -- 25 cases were detected at better than 5\% significance (a $2\sigma$ detection) when combining radial velocity measurements with two-dimensional spatial information, almost double the number of detections when using purely spatial substructure information. 

\citet{Cohn12}, using cosmological simulations, studied the velocity distributions of infalling subclusters and concluded that clusters are preferentially elongated along the infall directions of massive subclusters. \citet{Cohn12} showed using the Dressler-Shectman test (\citealt{Dressler88}, which uses spatial and radial velocity information to detect substructure) that the amount of detected substructure was uncorrelated with the line of sight used for detection in most clusters in their sample. Interestingly, their analysis also found that while cluster substructure was detected more often when it was perpendicular to the line of sight, $\sim 1/4$ of these clusters were more likely to be detected along lines of sight closer to the infall direction. 

The seemingly contradictory results of \citet{Cohn12} can be understood if one accounts for the fact that subclusters that fall in perpendicular to the line-of-sight do not have large radial velocity offsets from the main cluster, but can be detected spatially, while subclusters that fall in along the line-of-sight are not seen as being spatially distinct, but have large deviations in radial velocity and velocity dispersions from the main cluster. We explore this phenomenon in this paper. 

The positions and radial velocities of dwarf and giant cluster galaxies can be combined in the form of phase-space diagrams to gain further insights into the dynamical state of galaxy clusters. At a given radius, recently accreted galaxies have a higher velocity dispersion than older virialized cluster members (e.g., younger dE's in Virgo have higher velocity dispersions than older giants, as quantified in \citet{Conselice01} and to be discussed later in this paper). Additionally, a bound cluster's galaxies are confined to a characteristic trumpet-shaped `caustic' region in phase space, defined by the maximum escape velocity at a given radius (\citealt{Kaiser87}, \citealt{Regos89}, \citealt{Rines03}). In addition to having higher velocity dispersions, infalling and recently accreted galaxies can lie outside this caustic region, or escape velocity envelope, as we describe in this paper. In-depth photometric and spectroscopic studies of cluster galaxies, in particular their morphologies, star-formation rates, and velocities, are therefore crucial in probing the process by which clusters form and accrete their galaxies. 

In this paper, we describe the relationship between the velocity distribution of a cluster's galaxies and its dynamical state. In particular, we focus on the use of different galaxy populations' velocities as probes of their cluster's formation history, as well as the possibility of detecting the signature of an infalling group long after its first pericentric passage. We also describe the phase-space structure of a cluster that is in the process of accreting a massive subcluster, and the signatures of an infalling population in phase space, with a view towards using phase-space properties to detect otherwise indistinguishable phase-space structure. To accomplish these objectives, we perform a series of simulations of  group-cluster mergers in isolated boxes, with cosmologically consistent initial conditions, under the assumption that the group and cluster are collapsed systems whose evolution is largely unaffected by large-scale cosmic velocity fields.

The dynamical properties of infalling groups and the evolution of substructure in galaxy clusters can in principle be studied with cosmological simulations. We choose to use an idealized approach to quantify the unique effects of a merger on cluster dynamics, and the role of minor mergers in shaping the phase-space distribution of galaxy clusters. Using an idealized merger rather than a more realistic cosmological approach neglects multiple ongoing mergers of galaxies and small groups of galaxies, of various masses, along various directions. However, these effects are not as important to our current problem as that of the dynamics of a coherent bound group of galaxies. Although our clusters are spherically symmetric, which is not necessarily true for real cosmological clusters, the physical intuition and predicted results from our simulations are still useful for interpreting observations. For instance, earlier phase-space calculations based on cosmological simulations of clusters (\citealt{Serra11}, \citealt{Serra13}; see discussion in \S~\ref{sec:disc_phase}) are in reasonable agreement with models assuming spherical symmetry.  We plan to investigate the effects of infalling groups on cluster dynamics within a cosmological context in a future paper.

This paper is structured as follows: in \S~\ref{sec:methods} we describe our simulations' parameters and initial conditions. In \S~\ref{sec:results}, we illustrate the results of our simulations --- the orbital evolution of an infalling group, the evolution of group and cluster velocity dispersion particularly along a line of sight parallel to the merger, the evolution of velocity anisotropy, skewness, and kurtosis for the infalling group, and the group and cluster's evolution in phase space. In \S~\ref{sec:discussion} we discuss our results and compare them to observed clusters and their galaxies as well as other theoretical models of the dynamics of galaxy substructure. We summarize our results in \S~\ref{sec:summary}.

\section{Method}
\label{sec:methods}

The simulations in this paper were performed using \textsc{FLASH 4} (\citealt{Fryxell00}, \citealt{Dubey08}), a parallel $N$-body plus adaptive mesh refinement (AMR) Eurlerian hydrodynamics code. In  \textsc{FLASH 4}, particles are mapped to the mesh using cloud-in-cell (CIC) mapping, and a direct multigrid solver (\citealt{Ricker08}) is used to calculate the gravitational potential on the mesh. AMR is implemented using \textsc{PARAMESH 4} (\citealt{MacNeice00}). 

With our simulations, we explored a parameter space of group-cluster mergers to study the effect of group and cluster mass as well as group-cluster mass ratio on the velocity distribution of their post-merger components.  We performed a total of five $N$-body-only idealized simulations of group-cluster mergers, assuming standard $\Lambda$CDM parameters of $\Omega_{\Lambda} = 0.7$, $\Omega_{\rm m} = 0.3$, and $H_0 = 71 \, \mbox{km s}^{-1}\, \mbox{Mpc}^{-1}$. The cosmological parameters are used to calculate the mean density of the Universe and the redshift-dependent halo concentrations, scale densities, and $R_{200}$  radii (measured relative to the critical density), as further elaborated in \S~\ref{sec:ic}. The first four simulations were performed assuming a $z = 0$ critical density. To study the effect of increased density on our results, we performed the final simulation beginning at $z = 0.5$. The simulations performed are summarized in Table~\ref{tab:tab1}.

\begin{table*}
 \begin{center}
 \begin{tabular}{||c|c|c|c|c||}
  \hline
  Simulation & Cluster Mass & Group Mass & Density/Redshift   \\ \hline \hline
  M-2C-2G & $2 \times 10^{14}\, \mbox{M}_{\odot}$ & $2 \times 10^{13}\, \mbox{M}_{\odot}$ & Low-density, $z = 0$  \\
  M-2C-5G & $2 \times 10^{14}\, \mbox{M}_{\odot}$ & $5 \times 10^{13}\, \mbox{M}_{\odot}$ & Low-density, $z = 0$  \\
  M-5C-2G & $5 \times 10^{14}\, \mbox{M}_{\odot}$ & $2 \times 10^{13}\, \mbox{M}_{\odot}$ & Low-density, $z = 0$  \\
  M-5C-5G & $5 \times 10^{14}\, \mbox{M}_{\odot}$ & $5 \times 10^{13}\, \mbox{M}_{\odot}$ & Low-density, $z = 0$  \\
  M-5C-5G-highz & $5 \times 10^{14}\, \mbox{M}_{\odot}$ & $5 \times 10^{13}\, \mbox{M}_{\odot}$ & High-density, $z = 0.5$ \\ \hline
  
  \end{tabular}  
  \caption{Summary of simulation parameter values.} 
   \label{tab:tab1}
\end{center}

\end{table*}

We note here that since all of our particles have the same low mass ($10^8~ \mbox{M}_{\odot}$), we do not properly account for the effects of dynamical friction on orbits of individual galaxies, which will be most important for the most massive members of the system and thus have little impact on the dwarfs that are the topic of this study. Adopting the Chandrasekhar formula for the acceleration due to dynamical friction (\citealt{Binney08}, $a \propto M_{\rm particle}$), the acceleration due to gravity is $\mathcal{O} (10^3)$ times higher than the dynamical friction acceleration for a $10^9~ \mbox{M}_{\odot}$ test particle given our simulation setup, and therefore the effects of dynamical friction on dwarf galaxies can be safely neglected. 

\subsection{Initial conditions}
\label{sec:ic}
We used the cluster initialization technique of \citet{ZuHone11} and \citet{VR13} to initialize our group and cluster halos. The simulations described in this paper are pure $N$-body simulations with a uniform particle mass of $10^8~ \mbox{M}_{\odot}$. We briefly describe our halo initialization technique here, referring the reader to \citet{ZuHone11} and \citet{VR13} for a more detailed description. The group and cluster halos were initially assumed to be spherically symmetric, with total density profiles specified using a Navarro-Frenk-White profile (NFW, \citealt{Navarro97}) for $r \leq R_{200}$ with an exponential fall-off at $r > R_{200}$:
\begin{equation}
  \rho_{\rm tot}(r) = \begin{cases}
\displaystyle    \frac{\rho_{\rm s}}{r/r_{\rm s} (1 + r/r_{\rm s})^2} & {\hspace{-0.2in}} r \leq R_{200},\\
\displaystyle    \frac{\rho_{\rm s}}{c_{200}(1 + c_{200})^2}\left(\frac{r}{R_{200}}\right)^{\kappa} e^{-(r-R_{200})/{r_{\rm decay}}}   & \\
& {\hspace{-0.2in}} r > R_{200}.
  \end{cases}
\end{equation}
Here $r_{\rm decay} = 0.1R_{200}$, and $\kappa$ was chosen such that the density and the slope of the density profile are continuous at $R_{200}$. The concentrations, $c_{200} \equiv R_{200}/r_{\rm s}$, were derived from the redshift-dependent concentration-mass relationships of \citet{Duffy08}.

The parameters of the group and cluster halos used in our simulations are summarized in Table~\ref{tab:tab2}. We used the procedure outlined in \citet{Kazantzidis04} to initialize the positions and velocities of dark matter particles. For each particle, we drew a uniform random deviate $u$ in $[0,1)$ and chose the particle's halo-centric radius, $r$, by inverting the function $ u = {\int_0^r \rho (r) r^2 dr}/{\int_0^\infty \rho (r) r^2 dr}$.
To calculate particle velocities, we used the Eddington formula for the distribution function (\citealt{Eddington16}, \citealt{Binney08}):
\begin{equation}
  f(\mathcal{E}) = \frac{1}{\sqrt{8}\pi^2}\left[\int_0^{\mathcal{E}} \frac{d^2\rho}{d\Psi^2}\frac{d\Psi}{\sqrt{\mathcal{E} - \Psi}} + \frac{1}{\sqrt{\mathcal{E}}}\left(\frac{d\rho}{d\Psi}\right)_{\Psi=0} \right].
\end{equation}
Here, $\Psi = -\Phi$ is the relative potential of the particle and $\mathcal{E} = \Psi - \frac{1}{2}v^2$ is the relative energy. Using an acceptance-rejection technique, we chose random particle speeds $v$ given $f(\mathcal{E})$. 

\begin{table*}
\begin{center}
  \begin{tabular}{c c c c c c c c c c}
    \hline 
    Halo & $M_{200} (\mbox{M}_{\odot}$) & $z$ & $R_{200}$ (kpc) & $r_{\rm s}$ (kpc) & $\rho_{\rm s}$ ($\mbox{g}$ cm$^{-3}$) & $N_{\rm sat}$
    \\
    \hline
    \hline
    2G & $2 \times 10^{13}$ & 0 & 560.93 & 115.82 & $7.45 \times 10^{-26}$ & 24 \\
    5G & $5 \times 10^{13}$ & 0 & 761.3 & 169.77 & $6.26 \times 10^{-26}$ & 60 \\
    2C & $2 \times 10^{14}$ & 0 & 1208.3 & 302.77 & $4.83 \times 10^{-26}$ & 240 \\
    5C & $5 \times 10^{14}$ & 0 & 1640.17 & 443.8 & $4.08 \times 10^{-26}$ & 600 \\
    5G-highz & $5 \times 10^{13}$ & 0.5 & 636.32 & 171.69 & $7.03 \times 10^{-26}$ & 60 \\
    5C-highz & $5 \times 10^{14}$ & 0.5 & 1370.92 & 448.82 & $4.63 \times 10^{-26}$ & 600 \\
    \hline
  \end{tabular}
  \caption{Parameters of merging group and cluster halos. Note that $M_{200} = \frac{4}{3} \pi R_{200}^3 \times 200 \rho_{\rm crit}$. \label{tab:tab2}} 
\end{center}
\end{table*}

At the beginning of each merger, the separation between the group and cluster centers is $\Delta r = R_{200,{\rm g}} + R_{200,{\rm c}}$. The group infall velocities are given by $v_{\rm in} = 1.1 \sqrt{G M_{200,{\rm c}} / R_{200,{\rm c}}}$, consistent with the infall velocities derived from cosmological simulations in \citet{Vitvitska02} and used in \citet{ZuHone11}. Each merger was performed in a cubic box of side $13$ Mpc, with a minimum of 4 levels of refinement and a maximum of 7 levels of refinement, corresponding to a maximum spatial resolution of $12.7$ kpc. 

\section{Results}
\label{sec:results}
\subsection{Orbital Evolution}

\begin{figure}  
  \begin{center}
    {\includegraphics[width=0.5\textwidth]{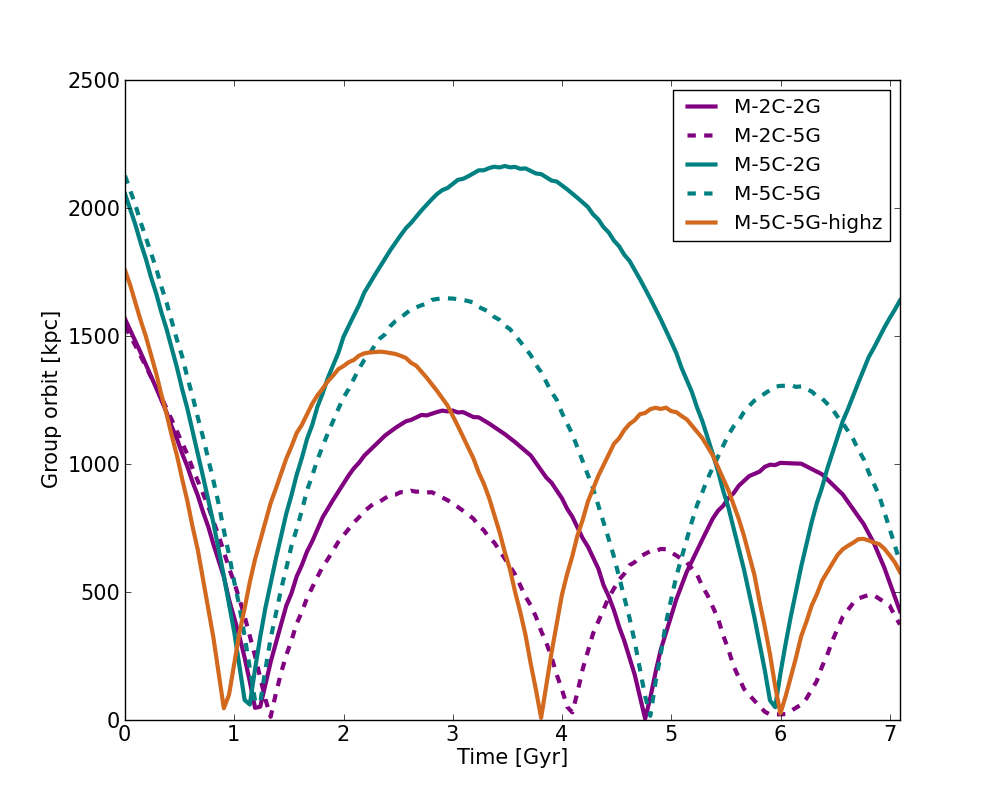}
    \caption{Evolution of the separation between the group's core and the system center of mass for all mergers.}
    \label{fig:orbit}}
  \end{center}  
\end{figure}

Figure~\ref{fig:orbit} shows the orbital evolution of the group in all five mergers. This plot shows the separation between the group's core and the combined system's center of mass as a function of time. For the four low-redshift mergers, the group's first pericentric passage is at $t \simeq 1.2 $ Gyr, and the high-redshift group's first pericentric passage is at $t \simeq 0.9 $ Gyr. We also see the effect of dynamical friction with different infalling group masses for a given cluster mass. Comparing the orbits of M-2C-2G with M-2C-5G, we see that the higher-mass group's orbit, after the first pericentric passage, has both shorter apocentric passage distances from the center of mass and a more rapidly decaying orbital period, compared to the lower-mass group. We see the same behavior when comparing the orbits of M-5C-5G and M-5C-2G, as well as M-5C-5G with M-5C-5G-highz. The increased density and consequently stronger dynamical friction in M-5C-5G-highz leads to smaller, more rapidly decaying orbits. 

\subsection{Velocity Distribution}
\label{sec:veldist}

\begin{figure*}
  \begin{center}
    \subfigure[Group, M-5C-5G]
    {\includegraphics[width=3.in]{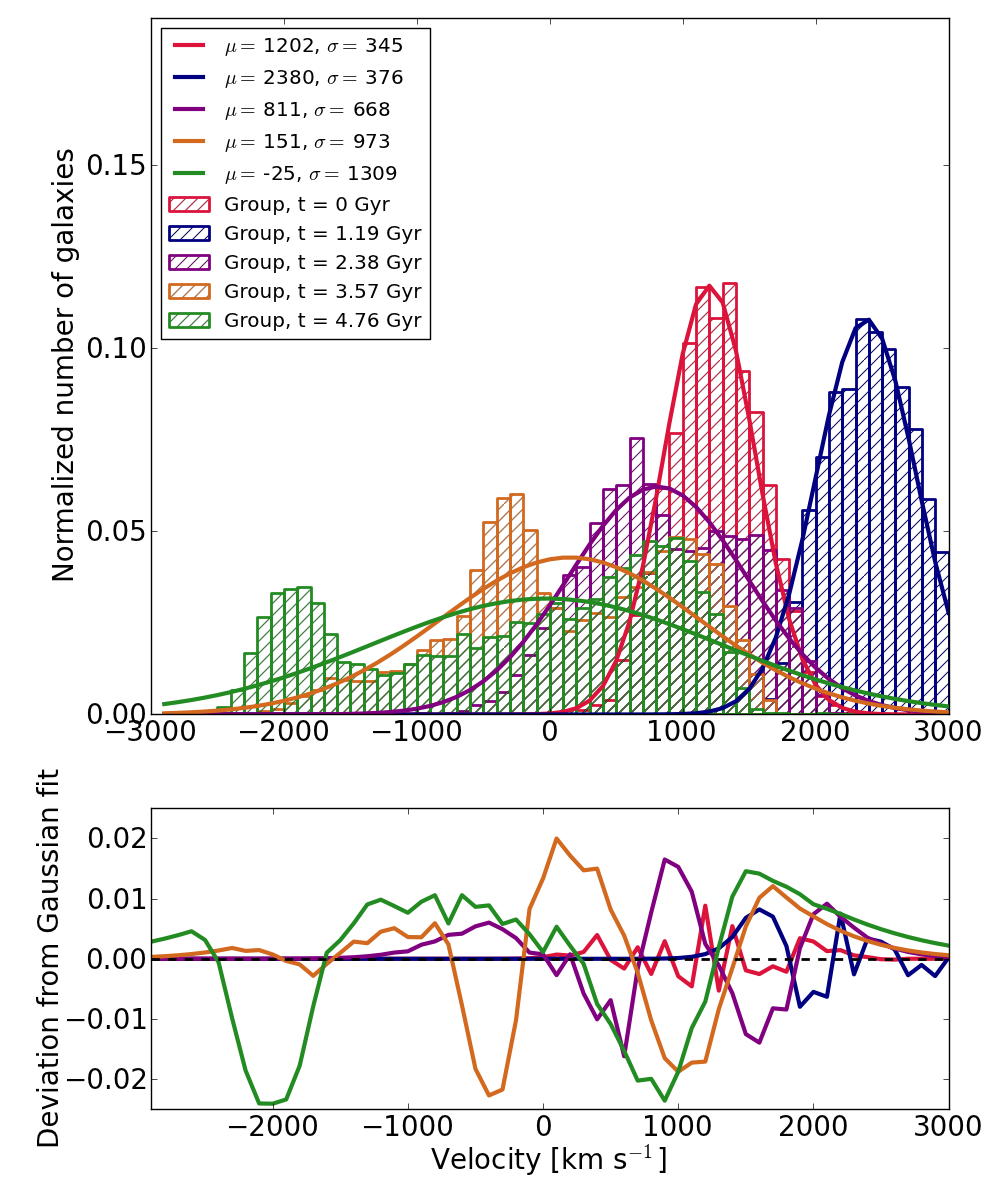}
    \label{fig:vel_dist_5c_5g_group}}
    \subfigure[Cluster, M-5C-5G]
    {\includegraphics[width=3.in]{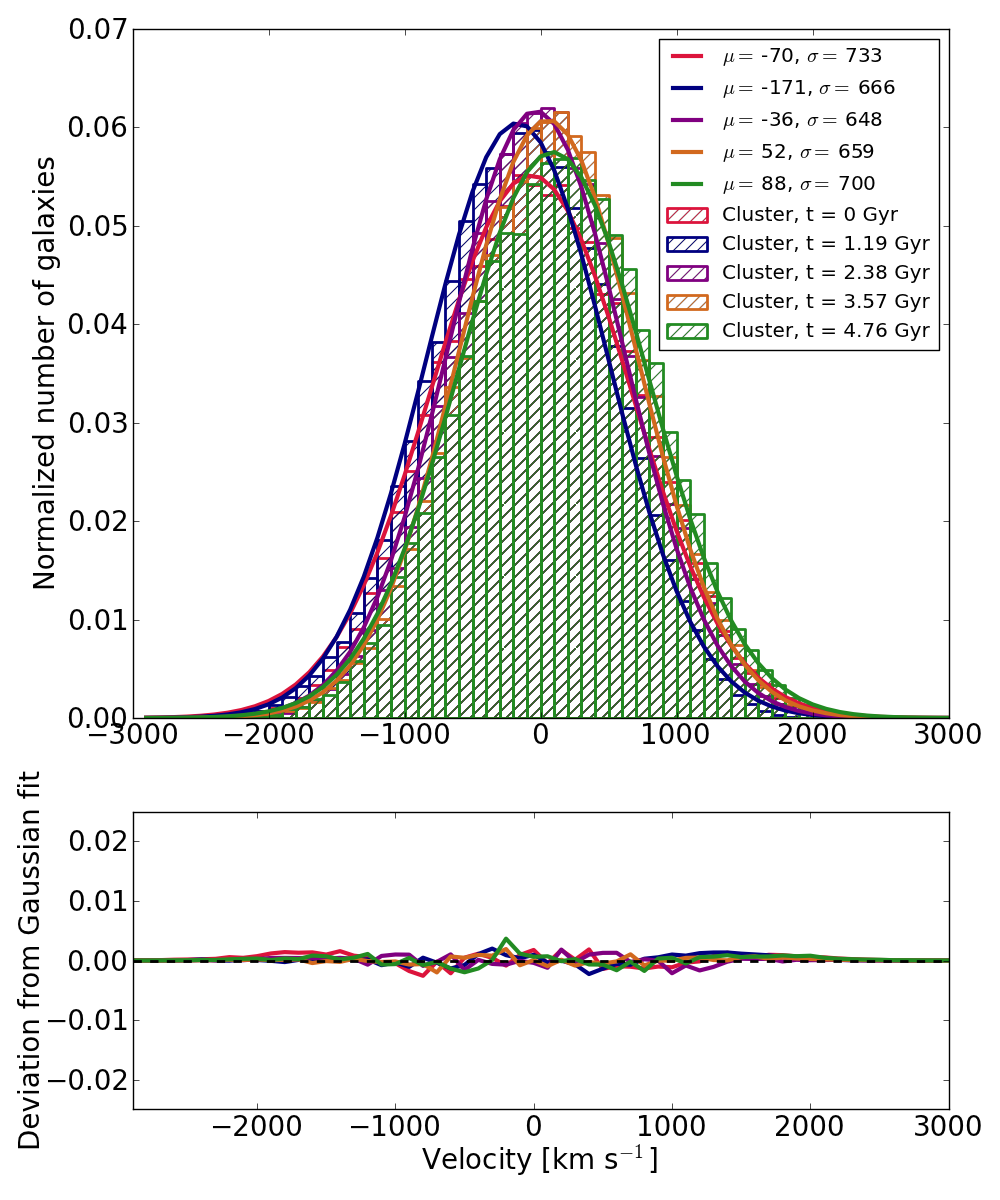}
    \label{fig:vel_dist_5c_5g_cluster}}
   \end{center}  
  
   \begin{center}
    \subfigure[Group, M-2C-5G]
    {\includegraphics[width=3.in]{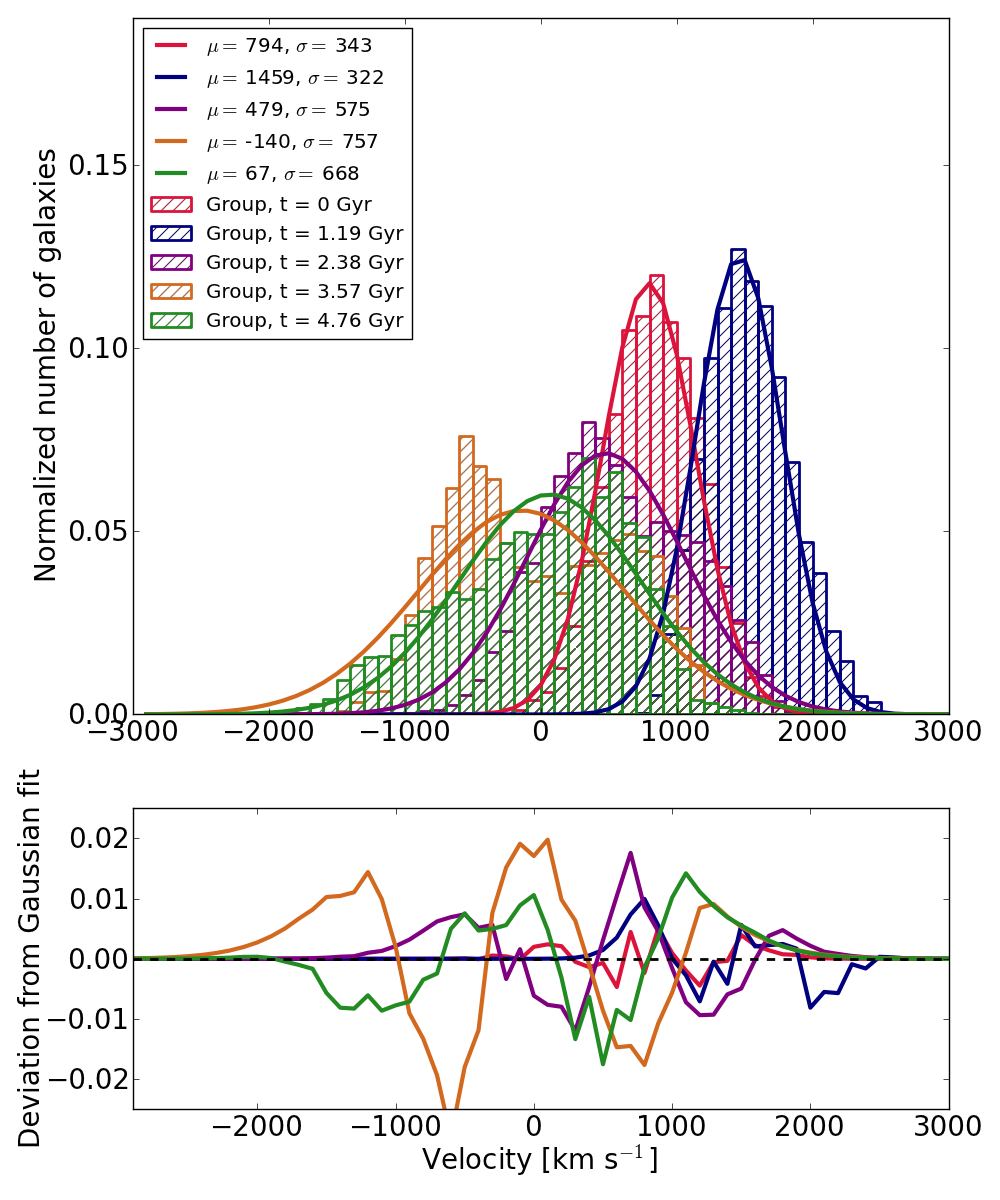}
    \label{fig:vel_dist_2c_5g_group}}
    \subfigure[Cluster, M-2C-5G]
    {\includegraphics[width=3.in]{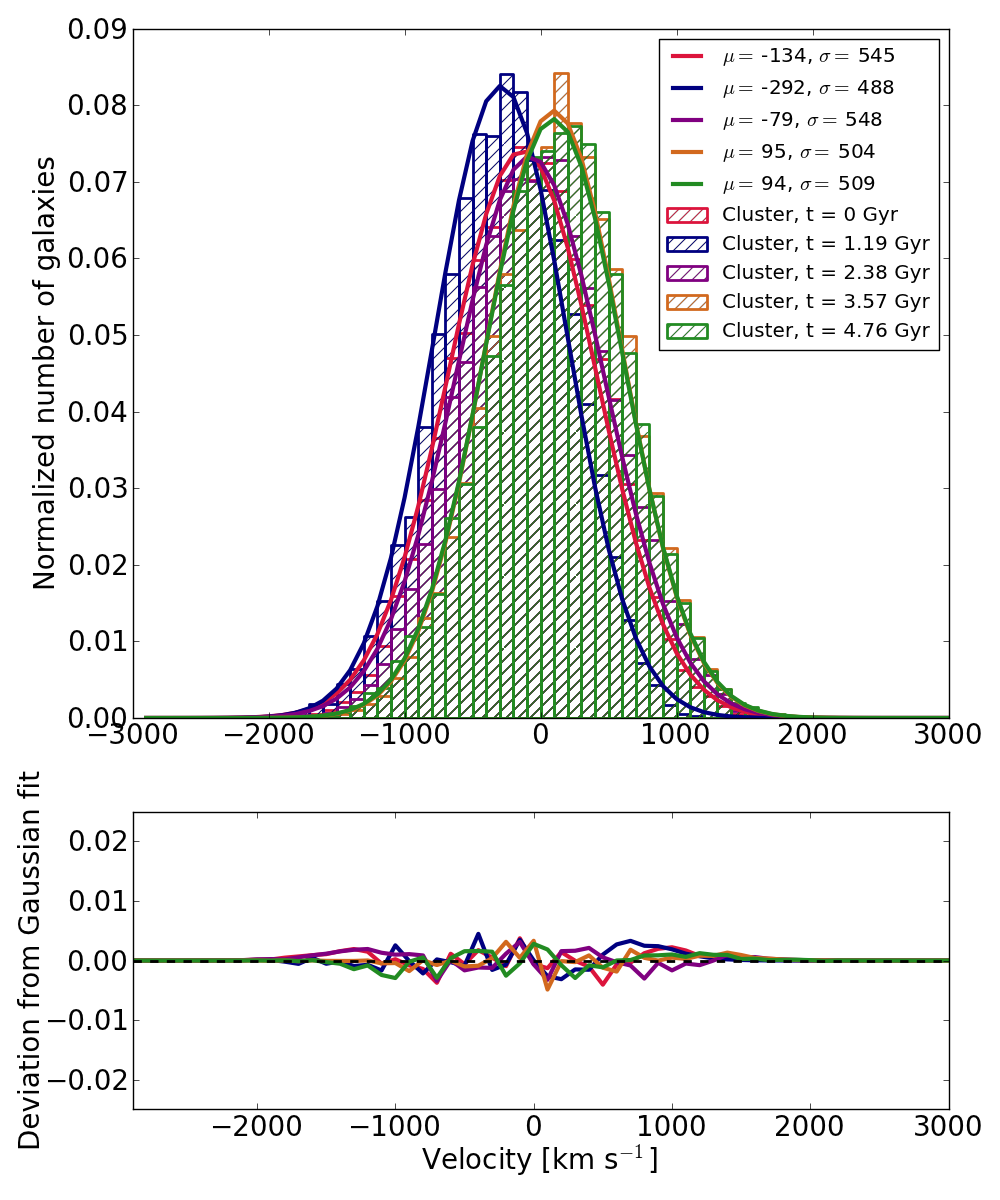}
    \label{fig:vel_dist_2c_5g_cluster}}
    \caption{Line of sight velocity histograms of the group and cluster, as viewed parallel to the infall direction. The solid lines correspond to the best-fit Gaussian profiles for each distribution, and the legend indicates the mean ($\mu$) and standard deviation ($\sigma$) of the Gaussian distribution in units of km s$^{-1}$.  These figures show the two most extreme-mass mergers: the upper panels show the velocity distribution in the $5 \times 10^{13}\, \mbox{M}_{\odot}$--$5 \times 10^{14}\, \mbox{M}_{\odot}$ merger, and lower panels show the $5 \times 10^{13}\, \mbox{M}_{\odot}$-$2 \times 10^{14}\, \mbox{M}_{\odot}$ merger. \label{fig:vel_dist}}
  \end{center}  
\end{figure*}

We use the velocity distributions of the cluster's and infalling group's particles to quantify the post-merger dynamics of cluster galaxies. We integrate the conditional mass function of \citet{Yang08} to estimate the number of galaxies more luminous than $10^8~\mbox{L}_{\odot}$ in our groups and clusters (Table~\ref{tab:tab2}). We then use a random ensemble of group and cluster particles' positions and velocities as proxies for galaxy positions and velocities. We stack 100 random realizations of `galaxy' particles to estimate the distribution of velocities. We note here that measured velocity distributions of observed cluster galaxies (e.g.\ \citealt{Conselice01}, \citealt{Drinkwater01}, \citealt{Lisker09}) are not as precise as those from our simulations. Additionally, we quantify the evolution of the merging group and cluster galaxies' velocity dispersions to physically motivate observed differences between the velocity distributions of different galaxy populations, although observations are restricted to measurements made at a single epoch.

Figure~\ref{fig:vel_dist} shows histograms of the group and cluster velocities in M-5C-5G and M-2C-5G. These plots show the velocity distribution as viewed along the direction of the merger, i.e., the direction of the group's infall is towards the observer. In both merger simulations, we see that the group's mean velocity is highest during the first pericentric passage ($\sim 1.2$ Gyr). After the pericentric passage, the group's mean velocity (with respect to the center of mass of the merged system) decreases. However, the group's velocity dispersion increases with time. At late times ($ t \gtrsim 3 $ Gyr), the group's velocity distribution is bimodal. The components in the bimodal distribution correspond to the group's core and its outer, less bound, rapidly stripped `halo' which is unbound soon after the group's first pericentric passage. The overall spread in group velocities along the merger direction remains higher than the cluster's velocity dispersion, consistent with the infalling group being unrelaxed along the direction of infall. Furthermore, the group in the higher mass ratio merger, M-5C-5G, has a larger velocity spread compared to the group in M-2C-5G, since the more massive cluster has a deeper potential well. Additionally, and unsurprisingly, the lower-mass cluster in M-2C-5G is more susceptible to dynamical disruption by the infalling group compared to the higher-mass cluster in M-5C-5G. This is  seen in Figures ~\ref{fig:vel_dist_2c_5g_cluster} and \ref{fig:vel_dist_5c_5g_cluster}, where the lower-mass cluster in Fig.~\ref{fig:vel_dist_2c_5g_cluster} has a higher mean velocity as well as a larger relative change in velocity dispersion during the pericentric passage. 

\subsubsection{Lines of sight and measured velocity dispersions}
\label{sec:angle}

\begin{figure*}
  \begin{center}
  \subfigure[M-5C-5G]
    {\includegraphics[width=3.2in]{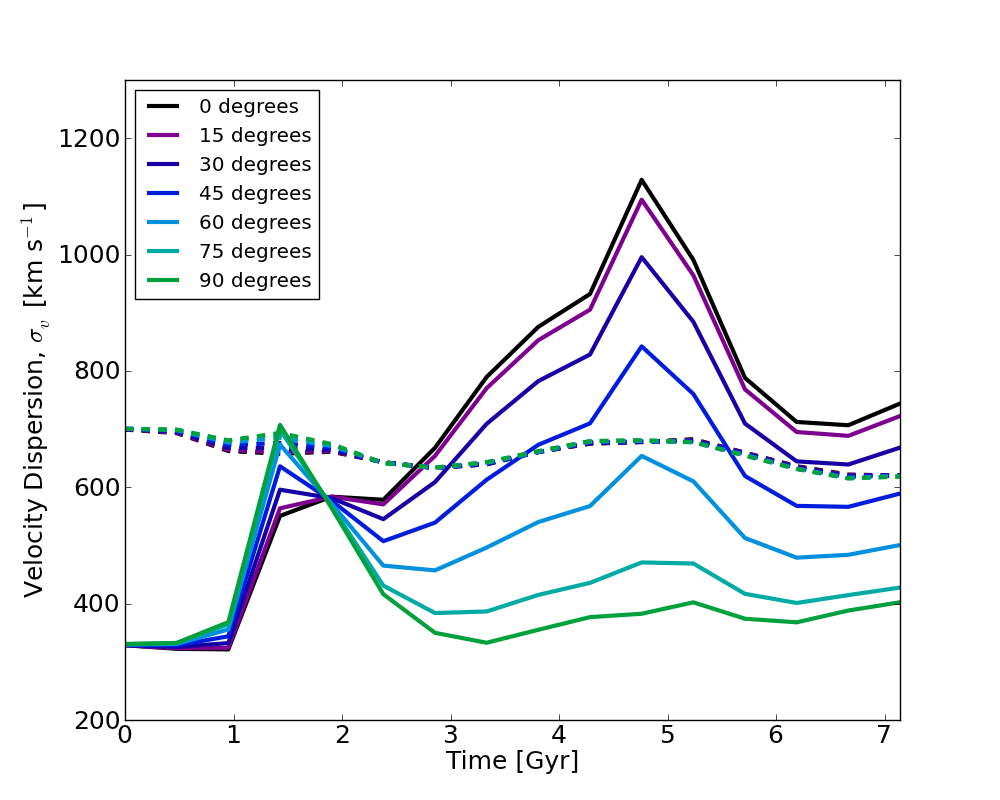}
    \label{fig:v_sd_5c_5g}}
   \subfigure[All mergers]
    {\includegraphics[width=3.2in]{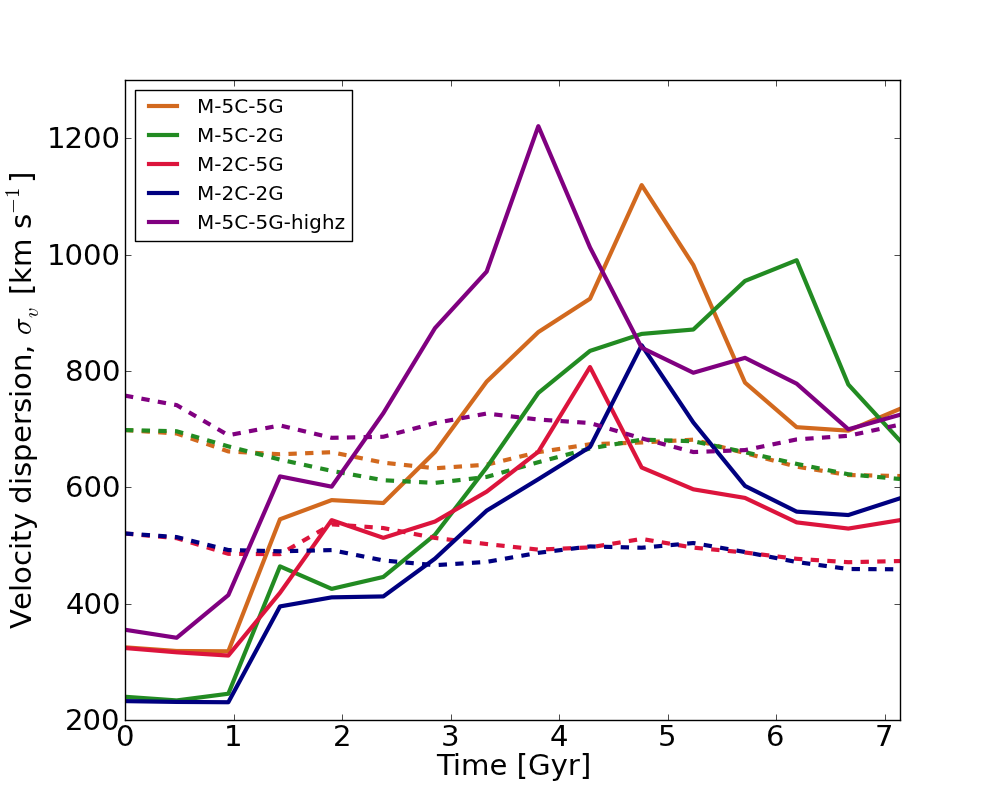}
    \label{fig:v_sd_all}}
    \caption{Left: The 1D line of sight velocity dispersions of the group and cluster viewed along different lines of sight for M-5C-5G. The dashed lines correspond to the cluster ($\sigma_{v, {\rm cluster}}$) and the solid lines to the group ($\sigma_{v,{\rm group}}$). The merger direction corresponds to the 0 degree lines, and the 90 degree lines correspond to direction perpendicular to the merger. Right: Line of sight velocity dispersions, along the infall direction, for all five mergers. \label{fig:v_sd}}
  \end{center}  
\end{figure*}

The one-dimensional velocity dispersion of the infalling group's components ($\sigma_{v, {\rm group}}$), and the ratio of the group galaxies' velocity dispersion to the cluster galaxies' ($\sigma_{v,{\rm group}}/\sigma_{v,{\rm cluster}}$) are functions of the viewing angle along which velocities are measured. Figure~\ref{fig:v_sd_5c_5g} shows the 1D velocity dispersions of the group and cluster's components in M-5C-5G, as measured along different lines of sight (indicated using different colors). There is an overall `heating' of the group during the pericentric passage, both parallel and perpendicular to the merger direction. During the second pericentric passage,  there is a significantly larger overall velocity boost along the merger direction than during the first passage. However, there is only a minor increase in $\sigma_{v,{\rm group}}$ perpendicular to the infall direction. Through the course of the merger, the group is `reheated' to the extent that for lines of sight that are within 45 degrees of the infall direction, the group's projected velocity dispersion is significantly higher than that of the cluster. This heating along the merger direction is a consequence of the decoupling in phase space of the group's core and halo components, an effect that is more apparent in Figure~\ref{fig:phase1d_par_5c_5g}, and is described in further detail in \S~\ref{sec:phase}. The cluster's velocity dispersion does not vary significantly as a result of the merger, except for a minor boost during the first pericentric passage.

\subsubsection{Variation in velocity dispersion with group and cluster mass}

\begin{figure*}
  \begin{center}  
      {\includegraphics[width=3.2in]{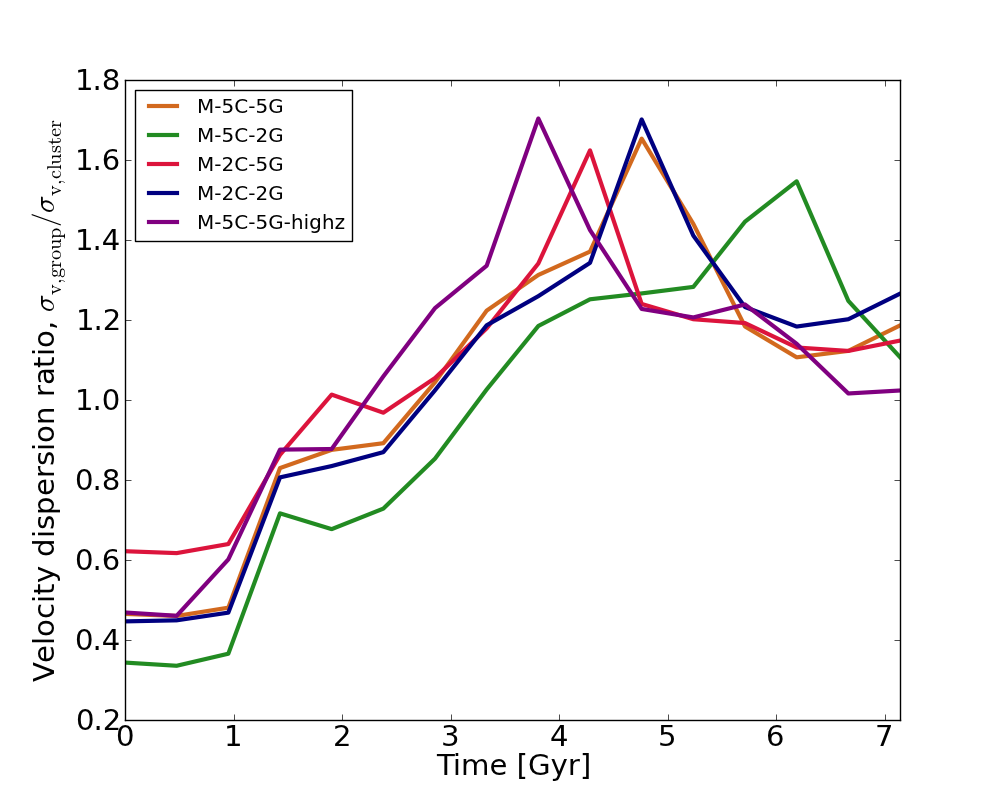}}
    	\caption{The ratio of the group galaxies' velocity dispersion to the cluster galaxies' velocity dispersion along the merger direction for all five mergers.\label{fig:v_sd_ratio}}
   \end{center}  
\end{figure*} 

In Figure~\ref{fig:v_sd_all}, we plot the line-of-sight velocity dispersions (along the merger direction) of group and cluster galaxies in all five mergers. Qualitatively, the evolution of velocity dispersion in the other four mergers resembles M-5C-5G (Fig.~\ref{fig:v_sd_5c_5g}). The group's velocity dispersion increases up to the first pericentric passage, briefly flattens, and then further increases up to the second pericentric passage, after which $\sigma_{v,{\rm group}}$ decreases to that of the cluster. For approximately two dynamical times \footnote{$t_{\rm dyn} \simeq (G \overline{\rho})^{-1/2}$} ($t_{\rm dyn} = 2.86$ Gyr at $z = 0$, 2.19 Gyr at $z = 0.5$), $\sigma_{v,{\rm group}}$ is 1.2 -- 1.8 times higher than $\sigma_{v,{\rm cluster}}$. This effect is more evident in Figure~\ref{fig:v_sd_ratio}, where we plot $\sigma_{v,{\rm group}}/\sigma_{v,{\rm cluster}}$ for all five mergers. Interestingly, the velocity dispersion ratio does not vary significantly between mergers of different masses and mass ratios. The maximum value of $\sigma_{v, {\rm group}}/\sigma_{v,{\rm cluster}}$ is $\sim 1.6  - 1.8$ and decreases to a value of $\sim 1.0 -1.2$ at the end of the simulation. 

The group-to-cluster velocity dispersion ratio is consistent with the infalling group galaxies forming an unvirialized population. The kinetic energy, $T$, and the potential energy, $U$, are related by $|T| \simeq 1/2 \, |U|$ for a population in virial equilibrium. However, as described in \citet{Colless96} and \citet{Conselice01}, for an accreted population $T + U \simeq 0$, so $|T| \simeq |U|$. Consequently, one expects that $\sigma_{v,{\rm infall}} \simeq \sqrt{2} \sigma_{v,{\rm virialized}}$. This is consistent with the velocity dispersion ratio of the infalling group to the virialized cluster seen in Figure~\ref{fig:v_sd_ratio}. We further compare these results to observed velocity dispersion ratios for real clusters in \S~\ref{sec:discussion_vdisp}.

\subsubsection{Velocity anisotropy}

\begin{figure*}
  \begin{center}  
      {\includegraphics[width=3.2in]{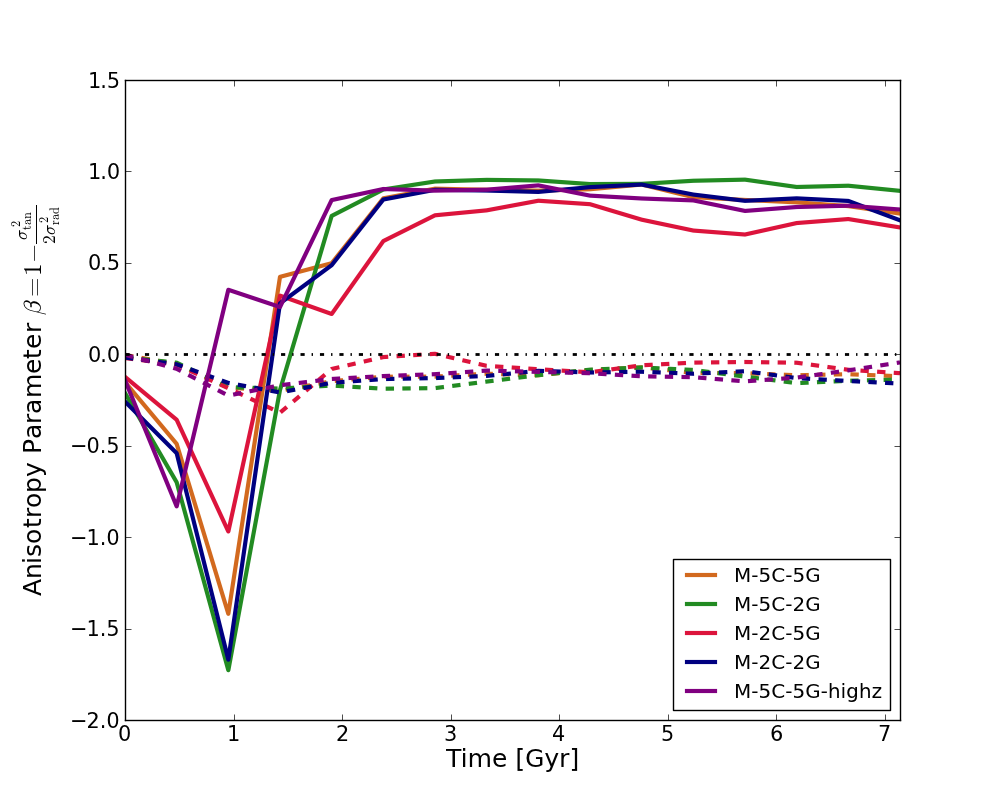}}
    	\caption{The anisotropy parameter, $\beta = 1 - \frac{\sigma_{\rm tan}^2}{2\sigma_{\rm rad}^2}$, of group and cluster galaxies. Solid lines correspond to group galaxies and dashed lines to cluster galaxies. The black dotted line indicates $\beta = 0$.\label{fig:v_beta}}
   \end{center}  
\end{figure*} 

As a consequence of the merger, the group and cluster galaxies' velocity distributions deviate from their initially assumed isotropy. The degree of anisotropy is quantified using the anisotropy parameter, defined as
\begin{equation}
\beta \equiv1 - \frac{\sigma_{\theta}^2 + \sigma_{\phi}^2}{2\sigma_{\rm rad}^2}   = 1 - \frac{\sigma_{\rm tan}^2}{2\sigma_{\rm rad}^2},
\end{equation}
where 
\begin{equation}
\sigma^2 = \overline{v^2} - \overline{v}^2 .
\end{equation}
For fully isotropic velocity dispersions, $\beta = 0$. For systems with radially biased orbits, $\beta > 0$, and for circular or tangentially biased orbits, $\beta < 0$. The deviations in the velocity distributions are calculated with respect to the mean center of mass velocity of the merging system, and the position vectors are measured with respect to the system's center of mass.

The evolution of the group and cluster galaxies' velocity anisotropies in all five mergers is seen in Figure~\ref{fig:v_beta}. The group and cluster are initially close to isotropic. During the pericentric passage ($\sim 1$ Gyr), the group and cluster galaxies' orbits are tangentially biased. However, after the pericentric passage, the group galaxies' velocities are highly radially biased, and the degree of radial anisotropy does not change significantly through the remainder of the merger. This persistence of anisotropy is consistent with earlier idealized and cosmological simulations of cluster formation. \citet{vanHaarlem93}, using a cosmological $N$-body simulation, showed that the presence of infalling substructure results in a higher radial to tangential velocity dispersion ratio. \citet{Roettiger97} used an idealized cluster merger approach and studied the evolution of velocity anisotropy. They showed that the radial bias in the velocity distribution of infalling substructure lasts for up to $\sim 5$ Gyr, consistent with our results. Consequently, one can conclude that radially biased velocity anisotropy is an signature of infalling substructure, but not necessarily recent infall. 

Unlike infalling group galaxies, the cluster galaxies' velocity anisotropies are not significantly affected by the merger. The degree of radial anisotropy of the group depends to some extent on the mass ratio of the merger. The group in the lowest mass ratio merger (M-2C-5G) has the lowest radial anisotropy, while the system with the largest mass ratio, M-5C-2G, has the highest radial anisotropy. The variation in $\beta$ for the cluster galaxies with mass ratio is not significant enough to indicate a trend.

\subsection{Detecting Infall Populations}
\label{sec:detect}
Observationally, detecting a subcluster whose infall direction is parallel to the line of sight (or LOS, the direction corresponding to the imaginary line connecting the observer to the cluster) is non-trivial. In this section, we describe the properties of the higher-order moments of the velocity distribution, skewness and kurtosis, during the merger process. We also describe the properties of the infalling group and cluster in LOS phase space.

\subsubsection{Skewness and Kurtosis}
\label{sec:skewnesskurtosis}

\begin{figure*}
  \begin{center}  
  \subfigure[M-5C-5G]
    {\includegraphics[width=3.2in]{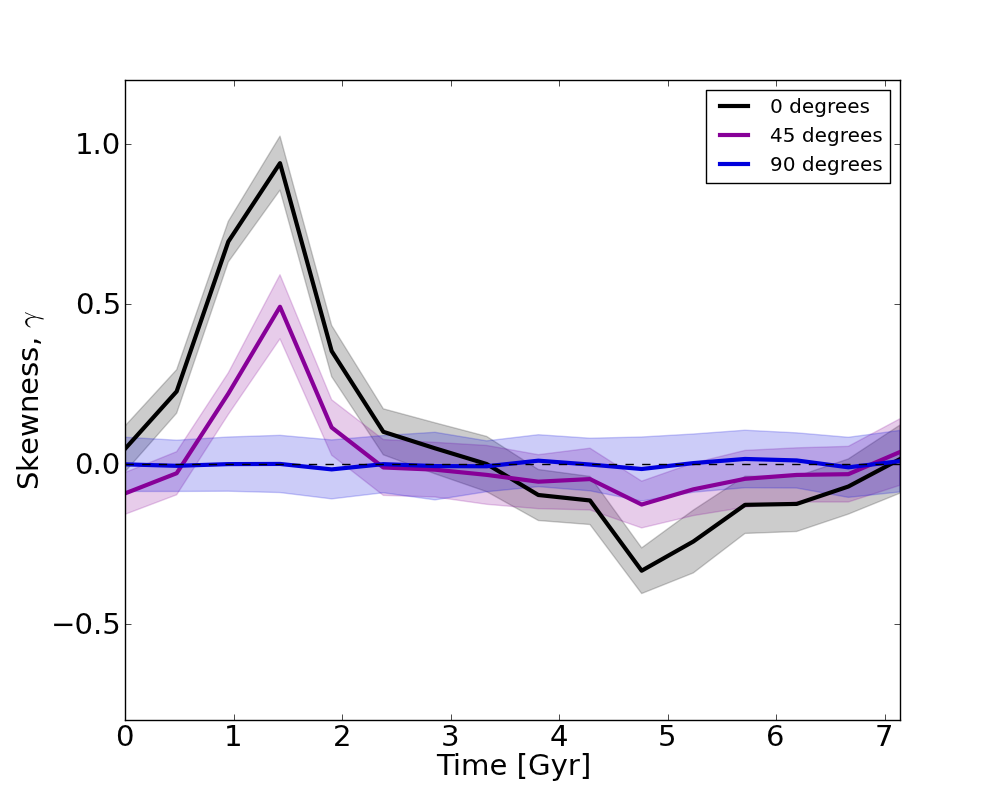}\label{fig:v_skew_5c_5g}}
  \subfigure[All mergers]
    {\includegraphics[width=3.2in]{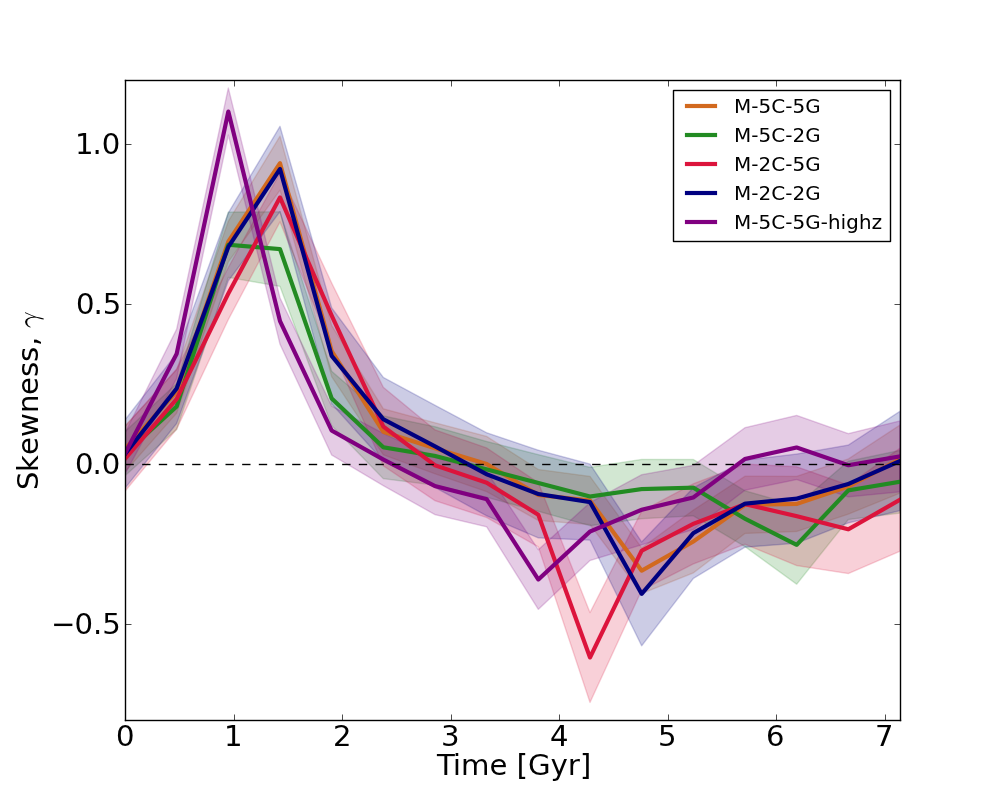}\label{fig:v_skew_all}}
    	\caption{Left: Skewness ($\gamma$) of the merged cluster's velocity distribution along varying lines of sight for M-5C-5G. The colors correspond to varying lines of sight with respect to the group's infall direction. The shaded regions correspond to the $1\sigma$ variation in measured skewness for the 100 random galaxy ensembles. Right: $\gamma$ along the line of sight parallel to the infall direction for all the merged clusters in our study. The black dashed lines correspond to $\gamma = 0$. \label{fig:v_skew}}
   \end{center}  
\end{figure*} 
 
\begin{figure*}
  \begin{center}
	\subfigure[M-5C-5G]	 
     {\includegraphics[width=3.2in]{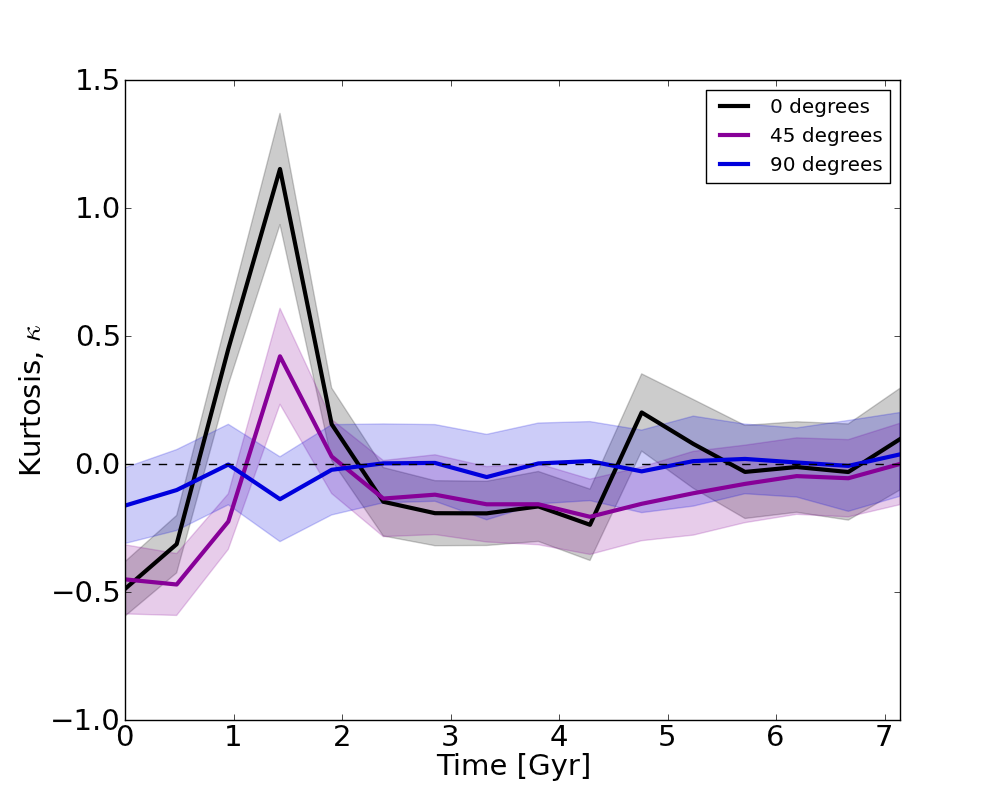}
     \label{fig:v_kurtosis_5c_5g}}
     \subfigure[All mergers]	    
     {\includegraphics[width=3.2in]{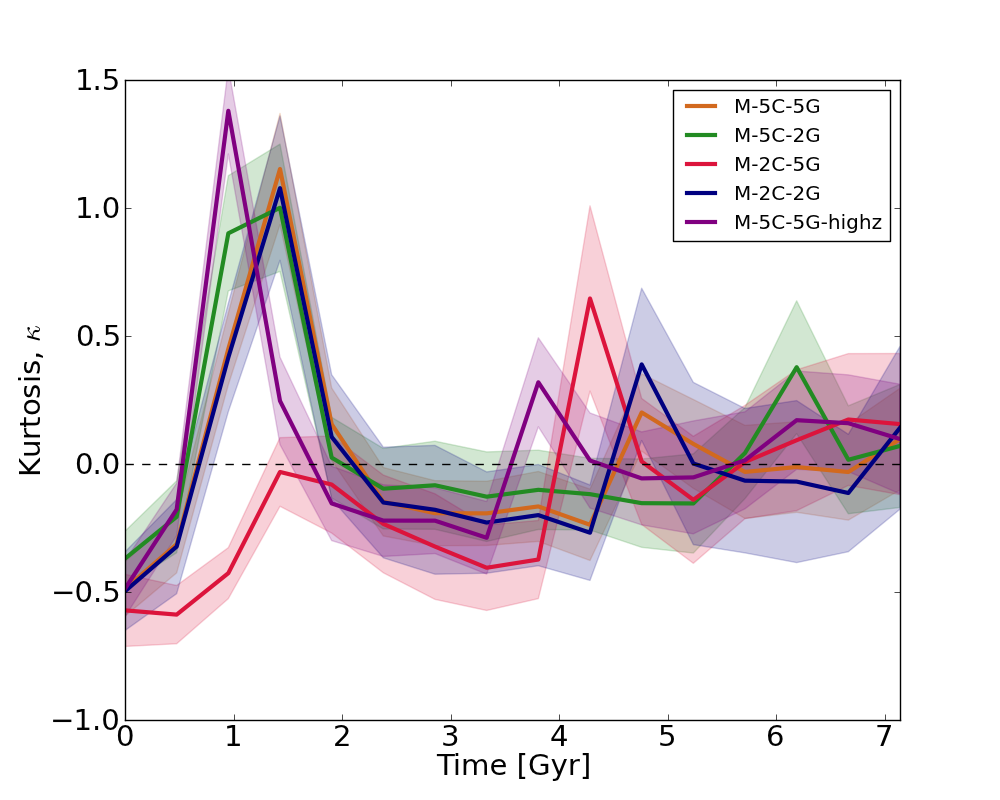}
     \label{fig:v_kurtosis_all}}    
    \caption{Left: Kurtosis ($\kappa$) of the merged cluster's velocity distribution for M-2C-5G and M-5C-2G, colors and shaded regions are as in Figure~\ref{fig:v_skew}. Right: $\kappa$ along the line of sight parallel to the infall direction for all the merged clusters in our study. The black dashed lines correspond to $\kappa = 0$.\label{fig:v_kurtosis}}
  \end{center}  
\end{figure*}

We quantify the deviation of the system's overall velocity distribution from a Gaussian using the skewness and kurtosis.  The skewness, $\gamma$, is defined as
\begin{equation}
 \gamma \equiv \frac{\langle(v_{\rm gal} - \overline{v})^3\rangle}{\sigma_{\rm v}^3}, 
\end{equation}
and the kurtosis, $\kappa$,  is defined as
\begin{equation}
 \kappa \equiv \frac{\langle(v_{\rm gal} - \overline{v})^4\rangle}{\sigma_{\rm v}^4} - 3. 
\end{equation}
The skewness is sensitive to the asymmetry of the distribution: $\gamma < 0$, or a negative skewness, corresponds to a longer left tail in a Gaussian distribution, and $\gamma > 0$ to a longer right tail. Kurtosis measures the `peakedness' of a distribution: $\kappa = 0$ corresponds to a Gaussian distribution, $\kappa > 0$ to a more peaked distribution, and $\kappa < 0$ to a flatter distribution.

Figure~\ref{fig:v_skew_5c_5g} shows the evolution of $\gamma$ as a function of time and viewing angle for the merging group-cluster system in M-5C-5G. We see a large positive skewness along the infall direction during the pericentric passage. This is a consequence of the group's net velocity boost in the direction of the merger. We also see a noticeable \textit{negative} skewness during the second pericentric passage, as the group travels in the opposite direction, away from the observer. As the group's core is accelerated during the second pericentric passage, the consequent velocity boost in the direction away from the observer results in the negative skewness. A smaller fraction of galaxies pass through the cluster core during the second pericentric passage compared to the first, so the magnitude of the skewness boost is comparatively lower. $\gamma$ is zero along the direction perpendicular to the merger, consistent with no velocity boost in this direction.  At intermediate angles along lines of sight within 45 degrees from the merger direction, $\gamma$ is non-zero during both pericentric passages. The shaded regions in this figure correspond to the  $1\sigma$ variation in measured skewness for the 100 random galaxy ensembles used in our calculation. Figure~\ref{fig:v_skew_all} shows the line-of-sight skewness for all five mergers, and we see the same qualitative behavior in all systems: a high positive skewness during the first pericentric passage, and a low negative skewness at the second pericentric passage. 

Figure~\ref{fig:v_kurtosis_5c_5g} shows the evolution of the system's kurtosis. The spikes in $\kappa$ at the two pericentric passages for smaller-angle lines of sight confirm that there is some compression in the directions parallel to the merger. This compression is significant for mergers within $45^{\circ}$ to the line of sight, based on the measured uncertainties. The overall evolution of the kurtosis along the infall direction varies with the mass of the group and cluster, unlike the skewness, as seen in Figure~\ref{fig:v_kurtosis_all}. The system with the smallest mass ratio (M-2C-5G) has the smallest kurtosis at the first pericentric passage and is the only merger in which the kurtosis further increases during the second pericentric passage. There also appears to be a net overall increase in the kurtosis of this system with time. In the other four systems, $\kappa$ is highest during the first pericentric passage, and the second peak, corresponding to the second pericentric passage, is lower than the first. This reflects a lower overall compression in velocity dispersion during the second pericentric passage. The compression during the first pericentric passage is driven by the compression along the infall direction in the \emph{cluster's} velocity distribution due to the group's core passage. This effect is more clearly seen in Figs.~\ref{fig:vel_dist_2c_5g_cluster} and \ref{fig:vel_dist_5c_5g_cluster}, where the cluster's $\sigma_{\rm v}$ decreases at $t = 1.19$ Gyr. As a result, the overall velocity distribution, including the group's high radial velocity components, is more peaked during pericentric passage. This effect is less pronounced during the second pericentric passage. 

The skewness and kurtosis are measured for the overall velocity distribution of the merged system, which is close to Gaussian. For certain mass ratios, the velocity distribution of the group alone is bimodal (Figure~\ref{fig:vel_dist_5c_5g_group}). Given the small number of group galaxies, the secondary peak in the group's bimodal velocity distribution (at $t = 4.76$ Gyr) does not make the overall velocity distribution bimodal. However, the velocity distribution of the merged system does deviate from Gaussianity, as reflected in the measurements of skewness and kurtosis at 4.76 Gyr. 

\subsubsection{Phase space structure}
\label{sec:phase}

\begin{figure*} 
  \begin{center}
    {\includegraphics[width=6.5in]{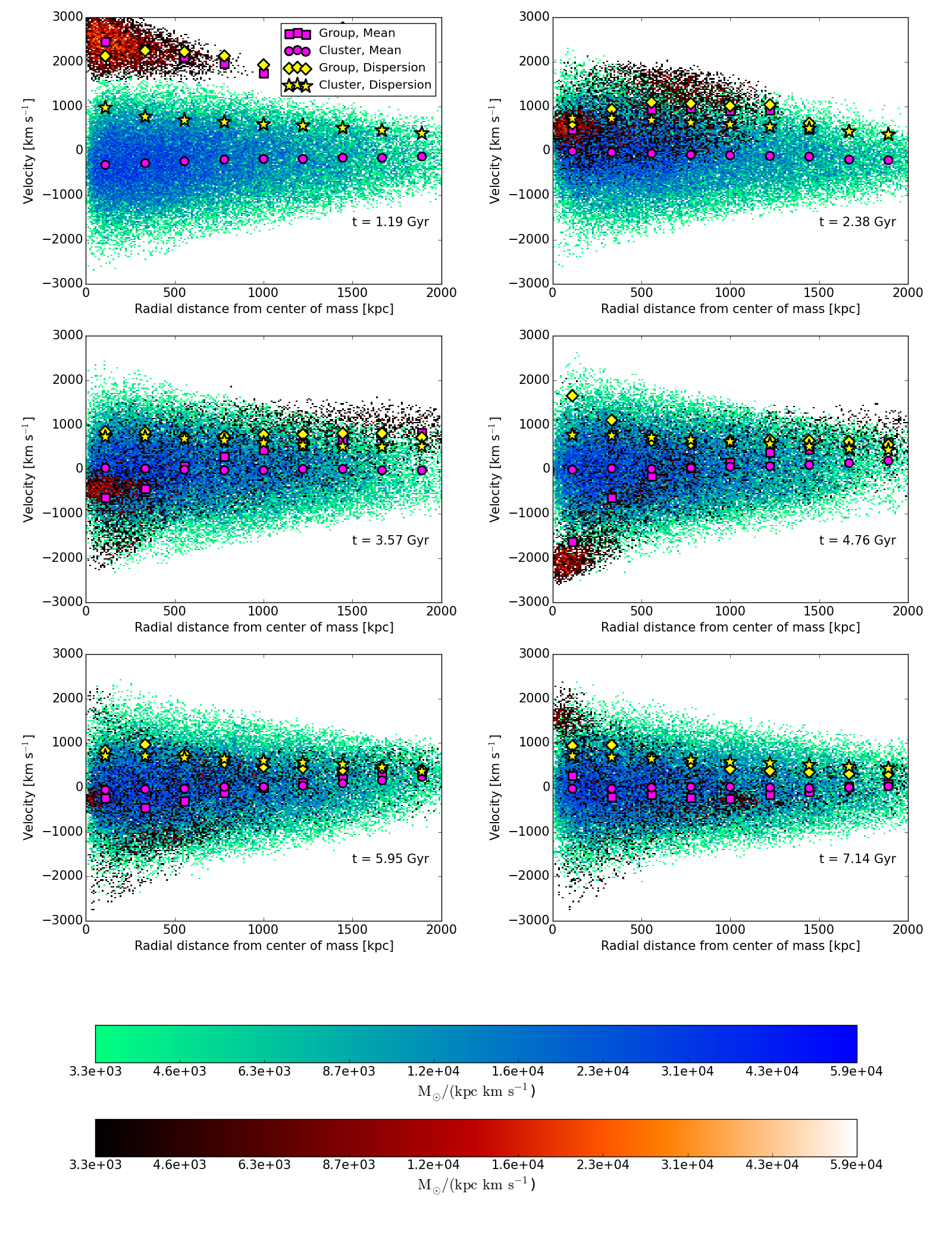}}
    \caption{Line-of-sight phase space density map of the group and cluster in M-5C-5G, measured parallel to the merger direction. Blue-green colors correspond to the cluster, and red-black colors to the group's components. The legend shows the two-dimensional phase-space density in units of $\mbox{M}_{\odot} /(\, \mbox{kpc} \, \mbox{km} \, \mbox{s}^{-1} )$. The orange symbols correspond to mean velocity in each cluster-centric radial bin for the group and cluster components, and the yellow symbols to the velocity dispersion.   \label{fig:phase1d_par_5c_5g}}
  \end{center}  
\end{figure*}

\begin{figure*} 
  \begin{center}
     {\includegraphics[width=7in]{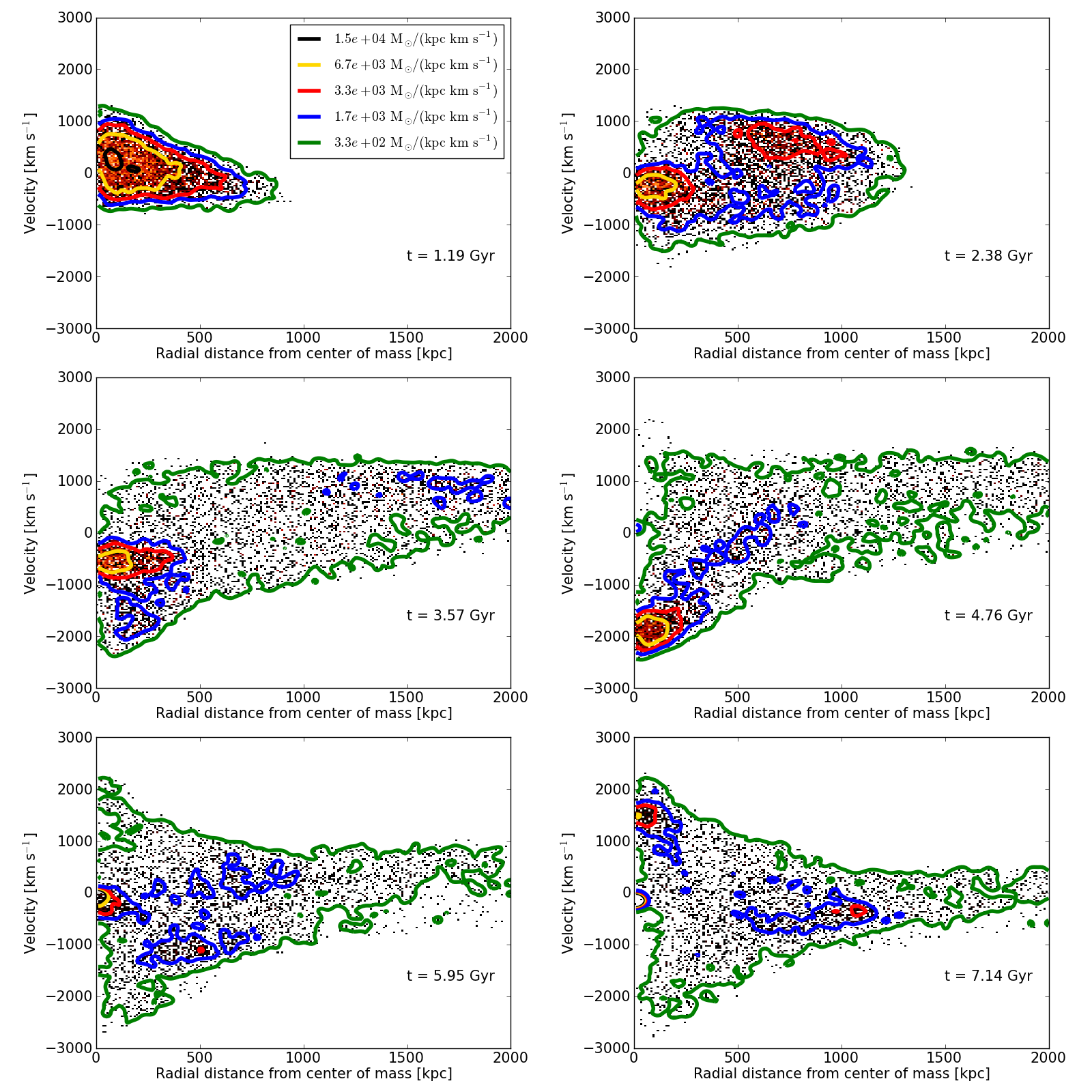}}
    \caption{Line-of-sight phase space density map of the group in M-5C-5G in the group's center-of-mass frame. Overplotted on the phase space map are contours of constant two-dimensional phase space density, in units of $\mbox{M}_{\odot} /(\, \mbox{kpc} \, \mbox{km} \, \mbox{s}^{-1} )$. \label{fig:phase1d_contour}}
  \end{center}  
\end{figure*}

\begin{figure*} 
  \begin{center}
    {\includegraphics[width=7in]{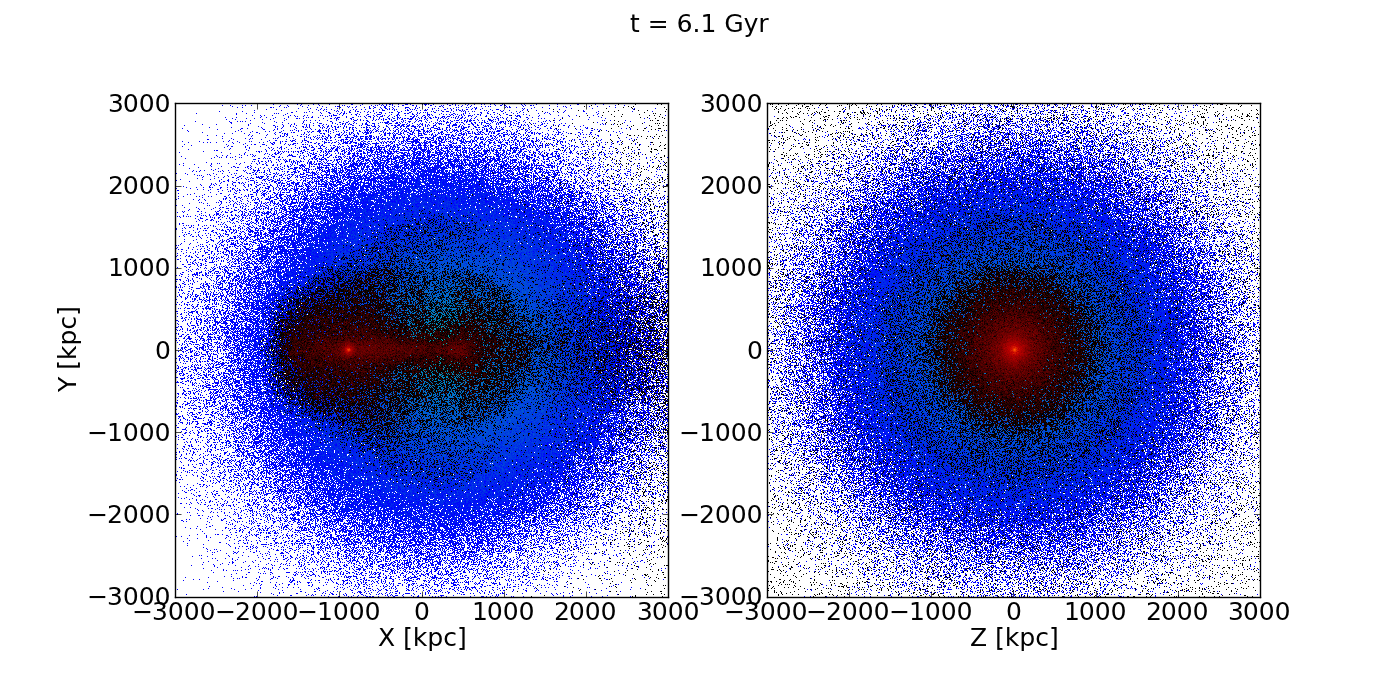}}
    \caption{Projected surface densities of the infalling group and cluster viewed perpendicular to (left) and along the direction of (right) the merger at the group's second apocentric passage in M-5C-5G.  \label{fig:gcdm_1_2_129}}
  \end{center}  
\end{figure*}

In this section we investigate the possibility of detecting signatures of an infall population in phase space for a merger along the line of sight. Figure~\ref{fig:phase1d_par_5c_5g} illustrates the evolution of group and cluster particles in phase space for M-5C-5G. We see the two distinct populations that correspond to the bimodality in the phase space diagram at $t = 2.38$, 3.57, and 4.76 Gyr in the phase space diagram. The infalling group's core region, whose components are within 500 kpc of the center, can be clearly distinguished in phase space. At later times, the infalling group's outer components tend to be located at large cluster-centric radii and also exist outside the main cluster's escape velocity envelope. This component is prominent at $t \simeq 3.5 - 5$ Gyr, before the group is eventually virialized within the cluster. As the group becomes bound to the cluster, its particles become restricted to the region within the cluster's escape velocity envelope.

We also calculate the velocity dispersion in each radial bin for both group and cluster components. We note here that the velocity dispersion in each radial bin is calculated with respect to the mean center-of-mass velocity of the group-cluster system, while $\sigma$ in Figure~\ref{fig:vel_dist} is the standard deviation of the best fit Gaussian to the overall velocity distribution of each population. The yellow symbols in Figure~\ref{fig:phase1d_par_5c_5g} correspond to the velocity dispersions in different radial bins.  Based on the group's velocity dispersion as a function of radius, we see that the group cools outside in: the group's velocity dispersion progressively increases with smaller cluster-centric radius, and the group's core remains hotter than the cluster.  Consequently, the overall relative heating (as described in Section~\ref{sec:veldist} and Figure~\ref{fig:v_sd_all}) of the group during the merger is primarily a consequence of the group's core remaining hotter than the cluster. 

To further study the destruction of the infalling group and its evolution in phase space, we plot the LOS phase space density (along the merger direction) of the group in M-5C-5G in the group's center-of-mass frame (Figure~\ref{fig:phase1d_contour}). The contours in this figure correspond to regions of constant phase-space density. The outermost envelope of the group in phase space, represented by the green contour, expands with time as the group spreads out in phase space. The inner red and yellow contours correspond to denser regions in phase space. As the group is tidally ripped apart and virialized, the dense center of the group shrinks; the innermost red contour, for instance, encompasses $\sim 600$ kpc at $t = 1.19$ Gyr and less than $100$ kpc at $t \gtrsim 5$ Gyr. Additionally, the outer green contour becomes less asymmetric about the velocity axis at late times ($t \gtrsim 5.9 $ Gyr) compared to $t = 3.5 - 5$ Gyr as a consequence of the group's virialization within the cluster. 

Furthermore, in Figure~\ref{fig:phase1d_contour}, we begin to see the two distinct populations in the phase space structure of the group's components beginning at $ t = 2.38 $ Gyr: the central core component, centered at $r = 0$, and the outer halo component. As the group settles within the cluster, the group core's projected distance from the cluster's center becomes small when observed from along the merger direction. However, it has a relative velocity of up to $2000$ km s$^{-1}$ with respect to the group center of mass. The other non-core component, on the other hand, has a mean radial distance of $\sim 1000$ kpc from the group's center of mass, with a maximum relative velocity of $1000$ km s $^{-1}$. This distinct outer halo component corresponds to a ring-like structure, visible at late times along lines of sight parallel to the merger direction (right panel, Figure~\ref{fig:gcdm_1_2_129}). The left panel of Fig.~\ref{fig:gcdm_1_2_129}, which shows the merger as viewed perpendicular to the merger direction,  also shows the distinct core component, which has been shaped by its predominantly radial orbit with the cluster in combination with dynamical friction, particularly near the cluster core.

\subsection{Mergers in the plane of the sky and the Perseus cluster}
\label{sec:perseus}

While the primary focus of this paper is disentangling the dynamics of infalling groups and their galaxies in line-of-sight group-cluster mergers, here we briefly note the primary characteristics of mergers in the plane of the sky. The Perseus Cluster is likely to have undergone such a merger. The spatial distribution of galaxies in Perseus is asymmetric (\citealt{Bahcall74}) and morphologically segregated (\citealt{Andreon94}, \citealt{Brunzendorf99}) with a higher fraction of spiral and irregular galaxies in the region offset from the elliptical-dominated cluster core. Additionally, X-ray observations (\citealt{Churazov03}) show that the gas temperature and surface brightness distributions are asymmetric and aligned with the galaxy asymmetry. Consequently, Perseus is likely in the process of merging with a subcluster in the plane of the sky, a resulted further supported by observations of large-scale gas motions (\citealt{Schwarz92}, \citealt{Simionescu12}).  

\begin{figure*} 
  \begin{center}
    {\includegraphics[width=7in]{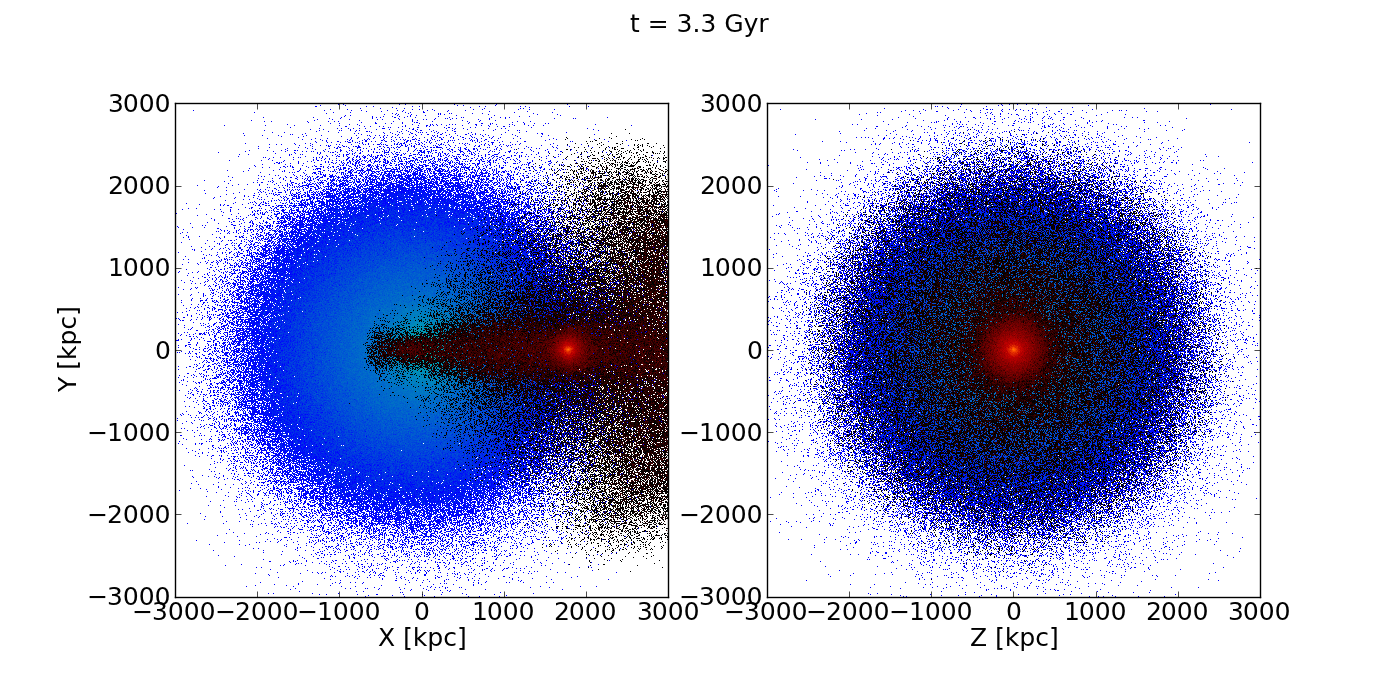}}
    \caption{Projected surface densities of the infalling group and cluster as viewed perpendicular to (left) and along the direction of (right) the merger at the group's first apocentric passage in M-5C-5G.  \label{fig:gcdm_1_2}}
  \end{center}  

  \begin{center}
    {\includegraphics[width=3.4in]{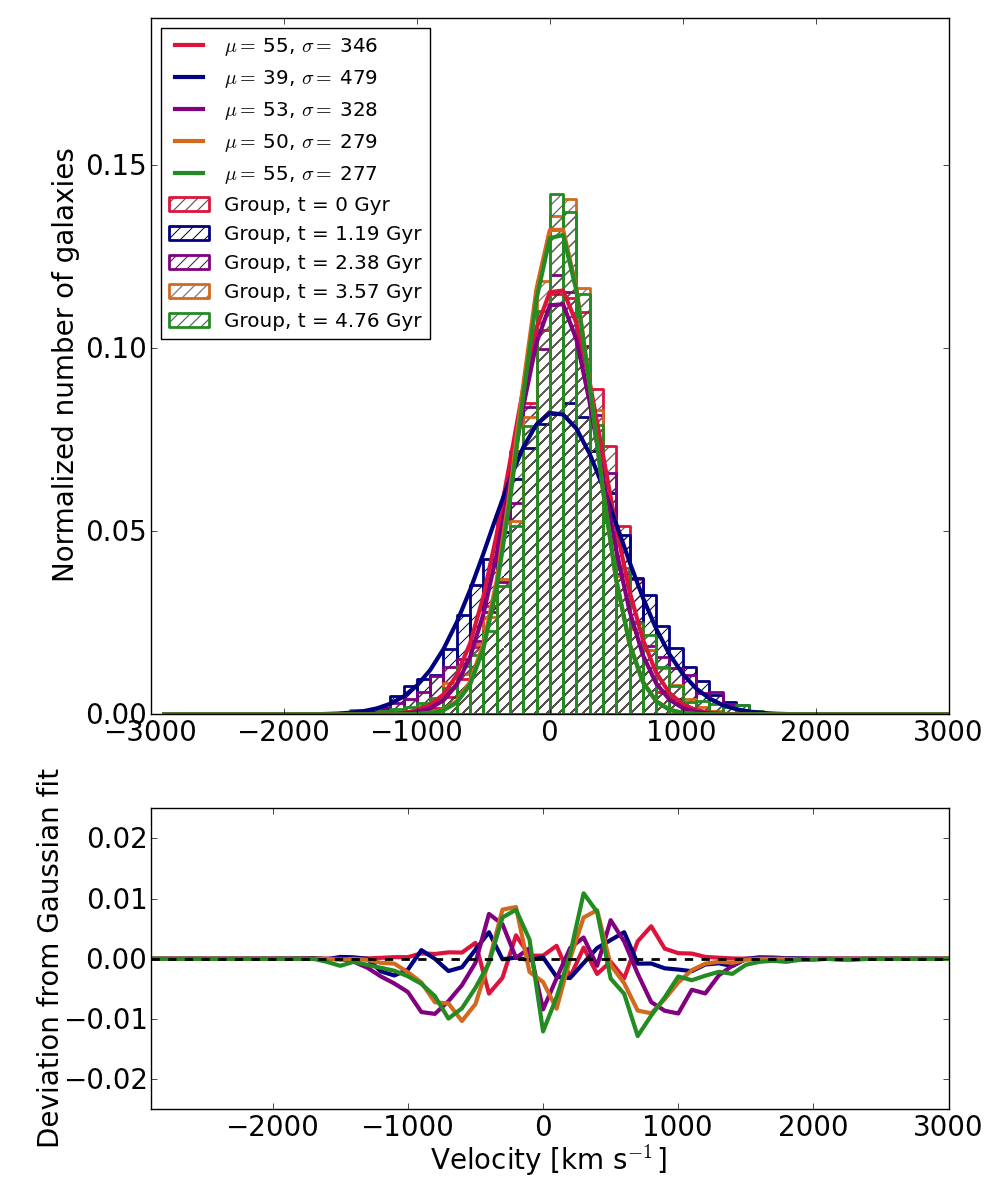}}
    \caption{The velocity distribution of the infalling group, as viewed perpendicular to the direction of infall, in M-5C-5G.  \label{fig:vdist_90}}
  \end{center}  
\end{figure*}

Our simulations, viewed perpendicular to the merger direction, are consistent with a Perseus-like scenario in which a younger group of galaxies falls into a relaxed cluster, and this group is currently near its orbital apocenter during the first infall. Figure~\ref{fig:gcdm_1_2} shows the group and cluster components projected in the directions perpendicular and parallel to the merger direction. The infalling group's components form an elongated structure along the infall direction. This spatial structure is qualitatively similar to the clustering of spiral galaxies in Perseus. Figure~\ref{fig:vdist_90} shows the velocity distribution of the group's galaxies during the merger. The group's velocity distribution does not change significantly as measured along the direction perpendicular to the merger; this is also seen in Figure~\ref{fig:v_sd}. Additionally, there is no net offset in the line-of-sight mean velocity of the group's galaxies.

\section{Discussion}
\label{sec:discussion}

Idealized simulations of cluster mergers are advantageous because they allow us to isolate and quantify the effects of a single merger, independent of the multiple ongoing mergers and accretion present in a cosmological simulation. By controlling the initial conditions, they enable us to explore the effect of merger parameters such as mass ratio and impact parameter without the trouble of locating appropriate merger events in a cosmological volume. While it is unlikely that any particular real group-cluster merger resembles in detail the simulated mergers in this paper, our simulations should nevertheless provide useful insight into the physics underlying real observations.

We interpret the results of our simulations in the context of the dynamics of cluster dwarf galaxies. The low masses of these galaxies and their insensitivity to dynamical friction make them excellent tracers of the merger history. As discussed in the introduction, the high dwarf to giant galaxy ratio in clusters compared to the field is likely a consequence of galaxy harassment and tidal stripping in the group and cluster environments. Dynamical friction can affect the orbital evolution of massive galaxies, and for a given galaxy mass, is more effective in lower mass clusters. The timescales over which dynamical friction acts are comparable to the Hubble time for only the most massive ($\sim 10^{12} \, \mbox{M}_{\odot}$) galaxies (based on Chandrasekhar's prescription for dynamical friction, as quantified in \citealt{Binney08}):
\begin{equation}
t_{\rm fric} = \frac{2.7 \, \mbox{Gyr}}{\ln \Lambda} \frac{r_{\rm inspiral}}{30 \, \mbox{kpc}} \left( \frac{\sigma_{\rm cluster}}{200 \, \mbox{km s}^{-1}} \right)^2 \left( \frac{100 \, \mbox{km s}^{-1}}{\sigma_{\rm satellite}} \right)^3 .
\end{equation} 
Here, $\sigma_{\rm cluster}$ and $\sigma_{\rm satellite}$ are the velocity dispersions of the cluster and satellite, $r_{\rm inspiral}$ is the initial radius of the satellite galaxy, and $\ln \Lambda \simeq 3$ is the Coulomb logarithm. Therefore, dynamical friction can be neglected for all but the most massive galaxies in relatively low-mass groups and clusters. 

\subsection{The evolution of the velocity distribution of infalling groups: Implications for detection}
\label{sec:discussion_vdisp}
Based on our simulations, we conclude that there exist two extreme scenarios for the visibility of infalling groups or subclusters. 

Subclusters that fall in parallel to the line of sight, in almost head-on mergers, are not spatially distinct, but can be distinguished in line-of-sight velocity space. These subclusters have high relative velocities with respect to the mean cluster velocity during the subcluster's infall and first orbital passage. At late times, the subcluster's mean velocity with respect to the cluster decreases. However, as the infall kinetic energy is transferred to the random motion of the subcluster's galaxies, the velocity dispersion of the infalling galaxies increases. The infalling subcluster's galaxies have a maximum velocity dispersion that is $1.6 - 1.8$ times higher than that of the cluster galaxies'. Our analysis of the velocity asymmetry of merging group-cluster systems, as measured by the skewness and kurtosis (Figs.~\ref{fig:v_skew} and ~\ref{fig:v_kurtosis}) show that overall velocity distribution is significantly non-Gaussian during the infalling system's pericentric passages.

Subclusters that fall in perpendicular to the line of sight are spatially distinct, forming extensions on one or two sides of the original cluster core, however, their galaxies cannot be distinguished in line-of-sight velocity space.  The measured mean velocities and velocity dispersions of these subclusters do not change significantly during the merger. 

One of the most convincing examples of the first scenario is the Virgo cluster.  \citet{Conselice01} quantified the dynamics of Virgo galaxies based on radial velocity measurements of 497 galaxies, including 142 dE + dS0 galaxies.  They showed that while giant elliptical galaxies are centrally concentrated with a Gaussian velocity distribution, indicating that they form a relaxed system, dwarf elliptical galaxies as well as spiral, irregular, and S0 galaxies are unrelaxed, less centrally concentrated populations. The velocity dispersions of late-type and dwarf galaxies are $\sim 1.5$ times higher than those of giant ellipticals. Later stellar age studies using population synthesis models by \citet{Lisker08} show that Virgo giants, in general, form an older population than the dwarfs. \citet{Lisker09} classified Virgo dEs based on their radial velocities and morphologies. They showed that flatter dEs are more likely to be on radial orbits, representing a more recently accreted population, while rounder dEs exist on more circularized orbits, representing an earlier generation of Virgo dwarfs. Other observational studies (\citealt{Sanchez12}, \citealt{DeLooze13}, \citealt{Rys14}) further support the hypothesis that a significant fraction of Virgo dwarfs are stripped, transformed populations, based on analyses of dwarf galaxies' globular clusters, dust-scaling relations, and circular velocity curves. The high velocity dispersions, diffuse spatial distribution, and morphological indicators of recent transformation therefore support a recent infall and transformation scenario for Virgo dwarfs.

The above observations of Virgo dwarfs suggest that the Virgo cluster has undergone one or more mergers, some possibly major, and the galaxies of the merged groups have possibly been subject to tidal stripping and harassment in their former group and present cluster environments. Our results show that infalling populations have radial velocity dispersions that are up to 1.5 times higher than virialized cluster populations and tend to exist at larger cluster-centric radii than pre-existing cluster populations. Additionally, the lack of any obvious spatial structure in the dEs suggests that Virgo must have undergone a head-on merger(s) along our line of sight. Our results also confirm that recently accreted galaxies tend to follow significantly more radial orbits, while older populations are on tangentially biased or circular orbits. 

While dwarf galaxies in general can be used to trace the overall dynamical history of a cluster, late-type galaxies form a separate, recently infalling galaxy population that is yet to be fully affected by galaxy transformation processes within the cluster. These populations therefore trace different stages of a cluster's dynamical evolution. Virgo dwarfs, which have a velocity distribution similar to late-type galaxies, show signatures of having been harassed or stripped. The dwarfs must have therefore spent a significant amount of time in a dense group or cluster environment in comparison to late-type galaxies to have been transformed into dwarf ellipticals. Taken together with their recent infall velocity signature, this implies that a significant number of Virgo dwarf galaxies have likely been pre-processed before being recently accreted into the Virgo cluster. 

Many dwarf galaxies in the Coma, Fornax, and Perseus clusters are consistent with being transformed late-type populations (\citealt{Graham03}, \citealt{DeRijcke03}, \citealt{Penny14}), based on observations of spiral arm remnants in two Coma dwarfs, embedded disks in Fornax dwarfs, and velocity dispersion measurements of Perseus dwarfs. Dwarfs and late-type galaxies in clusters often have systematically higher velocity dispersions than relaxed, older cluster populations, indicative of recent accretion of dwarfs. Giant elliptical galaxies are one such relaxed, older population (e.g.\ \citealt{Lisker08}). We note that because of their larger masses, the dynamical
friction timescales of giants are much shorter than for dwarfs (as shown for instance in \citealt{Jiang08}, \citealt{BoylanKolchin08}, and \citealt{Wetzel10}). Infalling giants are therefore more likely to merge with or be disrupted by the cluster central galaxy and overall potential. Additionally, their stars may have formed before they joined their current parent halos, so not all of the difference in stellar age and velocity dispersion can be reliably attributed to differing infall times. \citet{Drinkwater01}, using 108 galaxy velocity measurements of the Fornax Cluster, showed that the velocity dispersion of dwarf galaxies in Fornax is $\sim 1.4 $ times larger than that of the giants, consistent with a large fraction of dwarf galaxies being a recently accreted, yet to be virialized population. 

Fewer measurements of dwarf galaxy radial velocities in other clusters exist, hence we summarize existing measurements of late-type galaxy velocities that are consistent with their being recently accreted. \citet{Colless96}, based on radial velocity measurements of 465 Coma galaxies, found that the velocity dispersion of late-type galaxies is approximately $\sqrt{2}$ times that of early-type galaxies consistent with the idea that that they form a dynamically unrelaxed, recently accreted population. Early-type galaxies form the virialized cluster core in Coma. \citet{Ferrari03} found evidence for multiple Gaussian components in the velocity distribution of the Abell 521 cluster, including an infalling subcluster of predominantly late-type galaxies with a radial velocity of $\sim 3000$ km s$^{-1}$, and a velocity dispersion $\sim 1.5$ times higher than that of the overall cluster. Radial velocity measurements of dwarfs in these systems will help in further quantifying the dynamical state. 

\citet{Owers11} showed that Abell 2744, a merging cluster, has two distinct Gaussian velocity components corresponding to its merging subclusters. Interestingly, the galaxies identified as belonging to the smaller subcluster, based on their velocities, are spatially distributed in a central core region plus a less concentrated region spanning $\sim 2$ Mpc. This scenario in combination with X-ray data is consistent with  the subcluster being a post-core passage remnant. There is also evidence for significant correlation between galaxy morphologies and dynamics in galaxy cluster surveys and studies of ensembles of clusters. \citet{Biviano02}, using a sample of 59 clusters and 3056 cluster galaxies observed in the ESO Nearby Abell Cluster Survey (ENACS), showed that early-type galaxies are systematically more centrally concentrated with lower velocity dispersions than late-type galaxies. \citet{Biviano02} also showed that galaxies in identified subclusters have lower velocity dispersions than those outside subclusters.

Many observed clusters undergoing minor mergers exhibit skewed velocity distributions. \citet{Merritt87} and \citet{Fitchett87} showed that the Coma Cluster's galaxies have a skewed velocity distribution, suggesting that its dynamics are dominated by infall and accretion of galaxies. \citet{Bird94} studied a sample of 40 clusters with at least 50 measured redshifts each, and showed that the presence of substructure in clusters is correlated with the a significant measurable skewness and kurtosis. More recently, \citet{Mahajan13} showed that post-starburst (K+A) cluster galaxies are likely to be found in infalling subclusters and have positively skewed line-of-sight velocities, suggesting that K+A galaxies have been `pre-processed' and quenched of star formation in a smaller group environment before cluster infall. Skewness and kurtosis of galaxy velocity distributions are therefore good indicators of recent line-of-sight mergers, particularly during the core passage phase of a merger. 

Galaxy clusters in a cosmological context can have a non-zero kurtosis even in the absence of an active merger. Cosmological simulations (\citealt{Kazantzidis04}, \citealt{Wojtak05}, \citealt{Wojtak08}) show that the radial velocities of cluster dark matter particles within the scale radii of their host halos have a small positive kurtosis ($\kappa \simeq 0.5$), and particles in regions outside the scale radii have a negative kurtosis ($\kappa \simeq -0.5$). The high positive kurtosis ($\kappa \gtrsim 1$) in our simulated clusters during the merging groups' pericentric passages should therefore be measurable even in realistic cosmological clusters within the limits of uncertainty. 

The CLASH sample of clusters, along with follow-up VLT spectra (e.g.\ \citealt{Biviano13} \citealt{Annunziatella14}) can in principle be used to perform such an analysis with cluster dwarfs. However, the CLASH clusters have been explicitly X-ray selected to be relaxed, virialized clusters, and are therefore unlikely to exhibit signs of ongoing mergers. \citealt{Biviano13} find that the velocity dispersion profiles of blue galaxies are slightly higher than that of red galaxies, indicating that these galaxies have likely been recently accreted. A systematic analysis of the dwarf populations in these systems can provide further information on the accretion history of these systems. 

\subsection{Phase-space detection}
\label{sec:disc_phase}
Remnants of infalling groups, particularly in line-of-sight mergers, can be detected by combining velocity space information with spatial positions in phase-space diagrams. In these diagrams, the inner bound core and outer stripped halo are clearly visible as distinct components after the group's pericentric passage (Fig.~\ref{fig:phase1d_par_5c_5g}). The velocity dispersions of these components also evolve differently: the core has a lower velocity dispersion than the halo at early times ($t \lesssim 3.5$ Gyr). At late times, the group's core is disrupted by dynamical friction, and its components are dynamically heated during the group's orbital motion within the cluster ($t \gtrsim 4.7$ Gyr). By $7$ Gyr, the group's core cools as its velocity dispersion approaches that of the cluster.

To date, there exist only a handful of observational studies of cluster galaxies in phase space that include dwarfs, particularly in those clusters that have undergone head-on mergers. From the distribution of dwarfs and late-type galaxies in Virgo (Fig. 10 in \citealt{Conselice01}), we see that the majority of dE's in Virgo are likely remnants of an infalling group (or groups), and this group is still in the process of virializing its outer, more weakly bound halo. The velocity dispersions of the dE's and late-type galaxies in Virgo are higher than that of the giant ellipticals in all radial bins and do not fall off as a function of radius --- unlike the giant ellipticals, whose velocity dispersions do decrease with increasing cluster-centric radius. The radial dependence of the Virgo dEs' velocity dispersion, combined with the fact that their overall velocity dispersion is $\sim 1.5$ times higher than that of the giants, indicates that the infalling system from which some of the Virgo dE's originate is $\sim 1-2$ Gyr past its pericentric passage through the cluster. 

Caustics in redshift space have also been used to identify galaxies bound to clusters and measure cluster masses outside the virial radii (\citealt{Kaiser87}, \citealt{Regos89}, \citealt{Diaferio97}, \citealt{Geller99}). Caustics define boundaries of the escape velocity at a given radius, and for a bound system, the maximum velocity a particle or galaxy can have at that radius (\citealt{Diaferio97}, \citealt{Gifford13}). Galaxies that lie outside $R_{200}$ and are within the escape velocity envelope are in the process of falling into the cluster (\citealt{Geller11}). This region is referred to as the infall region. In redshift space, caustics take on a characteristic `trumpet' shape (\citealt{Kaiser87}, \citealt{Regos89}). Using cosmological simulations of clusters, \citet{Serra11} showed that the caustic technique recovers mass and escape velocity profiles on average with better than 10\% accuracy up to $4 R_{200}$. Clusters are in general assumed to be spherically symmetric in this technique. Deviations from spherical symmetry result in a 50\% uncertainty in individual cluster profiles. \citet{Serra13}, also using cosmological simulations, showed that this technique can identify cluster galaxies with a completeness fraction of 95\%. Observationally, galaxies outside these caustics are identified as not being bound to the cluster. \citet{Geller14} applied this technique to identify member galaxies of Abell 383. Among their results, they showed that blue galaxies did not affect the velocity dispersion within the cluster's virial radius, but at radii greater than $\sim 1 h^{-1}$ Mpc, the blue galaxies have a significantly higher velocity dispersion than the red galaxies. This is consistent with the bluer galaxies being a more recently accreted population that is in the process of being virialized.   

Identifying galaxy populations from infalling groups is the focus of our idealized simulations. We see that during the pericentric passage at $1.2$ Gyr (Figure~\ref{fig:phase1d_par_5c_5g}), the group's components lie well outside the cluster's escape velocity envelope since the group has a high infall radial velocity. During the merger, as the group becomes unbound and incorporated into the cluster, its components are increasingly confined to the region within the cluster's escape velocity envelope. Interestingly, at $\sim 3 - 5$ Gyr, the outer halo component of the group, at $r > 1000$ kpc, lies close to or outside the escape velocity envelope of the cluster. This feature can potentially be used to identify infalling galaxy populations a few Gyr after infall in real clusters. However, real clusters are often subject to multiple ongoing mergers, and their caustics include both members and non members due to projection effects, making the identification  and interpretation of individual subclusters complicated. In addition, real clusters are sparsely sampled (\citealt{Geller11}), making the identification of individual subcluster populations difficult. This problem can be mitigated with observations of dwarfs. Cosmological simulations (\citealt{Serra11}) show that a few tens of redshifts per square comoving megaparsec are sufficient to recover escape velocity profiles with the caustic technique. Radial velocity measurements of dwarf galaxies, which are more numerous than giant galaxies, can aid in extending this analysis to measuring the presence of substructure. A phase-space analysis of core and halo regions of infalling groups, correlated with measurements of the velocity dispersion of multiple cluster galaxy populations, will significantly improve our understanding of cluster formation histories through the detection of post core-passage substructure. 

\subsection{Interpretation of our results in the context of existing substructure detection tests}

The purpose of this paper is not necessarily to propose a specific or ideal substructure detection test, but to provide physical insight based on idealized simulated group-cluster mergers for the observed dynamics of cluster and subcluster galaxies, particularly dwarf galaxies. We relate the dynamical state of dwarf galaxies to the infall histories of their former hosts and show that spatial and kinematic signatures of infalling groups are not simultaneously detected for unfavorable lines of sight, particularly extreme cases of mergers parallel and perpendicular to the line of sight. 

\citet{Pinkney96} performed a comprehensive study comparing the effectiveness of various substructure detection tests that exist in the literature on simulated cluster mergers. They compared the effectiveness of 1D radial velocity based tests, 2D spatial distribution tests (including the symmetry and angular separation tests from \citealt{West88} and the Lee statistic from \citealt{Fitchett87}), and 3D spatial distribution plus radial velocity tests (including the \citet{Dressler88} test, the \citet{Bird93} test, and the \citet{West90} test). The conclusions from our models are consistent with their results. Their and our results show that 1D radial velocity tests are most sensitive in detecting substructure in line-of-sight mergers, particularly during core passage when there is no spatial substructure. These simulations also demonstrate that measured velocity dispersions increase significantly during the core passage, particularly for line-of-sight mergers, which additionally have high peculiar velocities. The models further indicate that two-dimensional spatial distribution tests can detect substructure in perpendicular mergers which do not display obvious velocity deviations, as seen in \S~\ref{sec:perseus}. Furthermore, their results show that 3D tests are most sensitive at detecting mergers that are $45^{\circ} - 60^{\circ}$ to the line of sight.  

\citet{Hou09} performed a more recent comparison of tests (including the $\chi^2$ test, the Kolmogorov test, and the Anderson-Darling test) designed to estimate the deviation of galaxy groups' velocity distribution from a Gaussian distribution. While they do not explicitly study the detection of substructure, their results are useful in broadly classifying the dynamics of groups. They show that dynamically unrelaxed groups with non-Gaussian velocity distributions have velocity dispersion profiles that increase with group-centric radius, while the opposite trend is seen in systems with Gaussian distributions, indicating that the former are dynamically unrelaxed systems. Our results, as illustrated in the phase-space diagram (Figure~\ref{fig:phase1d_par_5c_5g}), are consistent with this scenario. At early times ($t = 2.4$ Gyr), the infalling merging group's velocity dispersion increases with cluster-centric radius while the cluster's velocity dispersion decreases with radius.

\citet{Cohn12} analyzed the likelihood of substructure detection in clusters based on a cosmological simulation, by applying the Dressler-Schectman (DS) test along 96 lines of sight for each cluster. They find that the DS test is not always successful in detecting substructure along perpendicular lines of sight. However, they also find that a decrease in viewing angle relative to the merger direction resulted in increased sensitivity to subcluster detection in roughly a quarter of clusters with only major mergers. This makes sense, in the light of our simulations, when accounting for the fact that perpendicular mergers do not significantly affect both the group-centric velocity distribution and the peculiar velocity of the infalling group, two metrics to which the DS test is sensitive. 

\section{Conclusions}
\label{sec:summary}

We have used a series of idealized head-on galaxy group-cluster mergers to interpret the observed dynamics of dwarf galaxies in galaxy clusters as remnants of infalling groups.  We calculate the measured one-dimensional velocity dispersion of the infalling group for a range of merger mass ratios and viewing angles. We find that head-on mergers that are parallel to the observer's line of sight result in large radial velocity boosts during core passage, and an increase in velocity dispersion for the groups' galaxies as the merged groups are reheated during their passage through the clusters' potential well. This effect is noticeable in velocity space for mergers along lines of sight up to 45 degrees. As measured along the merger, the infalling groups have velocity dispersions that are up to $1.6 - 1.8$ times higher than that of the cluster, since the group's remnants are an unrelaxed population. The velocity distributions of these infalling systems are also radially biased after pericentric passage, and the skewness and kurtosis of their velocity distributions show a deviation from purely Gaussian distributions during pericentric passages. We also show, consistent with the results of \citet{Pinkney96}, that mergers parallel and perpendicular to the line of sight look qualitatively and quantitatively  different in their velocity structure and spatial distribution, and a single test based on velocities or positions alone cannot accurately identify both types of mergers.

These kinematic results for head-on mergers are consistent with measurements of the dynamics of dwarf galaxies in a few clusters (\citealt{Binggeli93}, \citealt{Conselice01}, \citealt{Drinkwater01}, \citealt{Lisker09}), which show that a significant fraction of cluster dwarfs have higher velocity dispersions than the clusters' giant ellipticals. The dynamics of these cluster dwarf galaxies can be explained if one considers that these cluster dwarfs are remnants of a merged group or subcluster, and additional `pre-processing' by the group has contributed to transforming former spiral galaxies into dwarfs in pre-merger group environments. Groups can contribute to transforming cluster galaxies through increased galaxy-galaxy merger rates and stripping of gas (e.g., \citealt{VR13}),  and a significant fraction of the Virgo Cluster's galaxies, particularly at large cluster-centric radii, are consistent with being pre-processed (\citealt{Gallagher89}, \citealt{Boselli06}, \citealt{Sanchez12}, \citealt{DeLooze13}, \citealt{Rys14}). We predict that future high-resolution spectroscopic observations of a large sample of cluster galaxies including dwarfs will be able to further quantify and detect the remnants of line-of-sight merged groups.  

In addition to purely kinematic detections, the phase-space structure of clusters with infalling groups, even along lines of sight parallel to the infall direction, can be useful in detecting and quantifying the extent of substructure. The infalling system's concentrated central core and diffuse halo can be distinguished in phase space (Figures~\ref{fig:phase1d_par_5c_5g} and ~\ref{fig:phase1d_contour}). The halo's remnants are found at larger cluster-centric radii at late times, outside the cluster's escape velocity envelope, and they have velocity dispersions higher than that of the cluster. The core appears compact in phase space and survives as a distinct component until dynamical friction destroys its coherence during repeated pericentric passages. Additionally, our results (Figure~\ref{fig:phase1d_par_5c_5g}) show that the caustic structure of the cluster is relatively insensitive to the merger and remains a good diagnostic to identify the cluster's escape velocity envelope. 

Clearly, there is a need for many more sensitive high-resolution spectroscopic observations to quantify the dynamics of clusters and study their merger histories. We show that it is possible, using velocity information, to distinguish the remnants of line-of-sight mergers. Dwarf galaxies (which do not necessarily have to be transformed in the cluster since they could have been transformed in groups prior to cluster infall) have an important observational role to play in this context since they are the most populous galaxy type in clusters and therefore act as effective tracer particles of their cluster's dynamical history. 

\section*{Acknowledgments}

The simulations presented here were carried out using the NSF XSEDE Kraken system at the National Institute for Computational Sciences and the Stampede system at the Texas Advanced Computing Center under allocation TG-AST040034N. FLASH was developed largely by the DOE-supported ASC/Alliances Center for Astrophysical Thermonuclear Flashes at the University of Chicago. This work was partially supported by the Graduate College Stutzke Dissertation Completion Fellowship at the University of Illinois at Urbana-Champaign. JSG thanks donors to the University of Wisconsin-Madison College of Letters \& Science for partial support of this research. We are grateful to Margaret Geller for her careful reading of the manuscript and insightful suggestions. We thank Chris Miller for useful discussions. We also thank the anonymous referee for useful comments and suggestions that improved this paper.

\bibliography{ms}

\end{document}